\documentclass[11pt,a4paper,twoside,groupcitations]{article}
\pdfoutput=1
\usepackage{graphicx,epsf}
\usepackage{amsmath}
\usepackage[]{latexsym}
\usepackage{amsfonts}
\usepackage{array}
\usepackage{amsthm}
\usepackage{amssymb}
\usepackage{braket}
\usepackage{eucal}
\usepackage{verbatim}
\newcommand{\be}{\begin{eqnarray}}
\newcommand{\ee}{\end{eqnarray}}
\newcommand{\pro}[2]{\mbox{$\langle\, #1 \mid #2\,\rangle$}}
\newcommand{\expec}[1]{\mbox{$\langle\, #1\,\rangle$}}

\renewcommand{\d}{\mbox{${\rm d}$}} 
\newcommand{\lp}{\ell_{\rm p}}
\newcommand{\mpl}{m_{\rm p}}
\newcommand{\gn}{G_{\rm N}}
\newcommand{\rh}{r_{\rm H}}
\newcommand{\Rh}{R_{\rm H}}
\newcommand{\Ah}{A_{\rm H}}
\newcommand{\K}{\mathcal{K}}
\newcommand{\Th}{T_{\rm H}}

\title{\bf Thermal BEC black holes}
\author{Roberto~Casadio$^{ab}$\thanks{E-mail: casadio@bo.infn.it},
$\ $
Andrea~Giugno$^{ab}$\thanks{E-mail: andrea.giugno2@unibo.it},
$\ $
Octavian~Micu$^{c}$\thanks{E-mail: octavian.micu@spacescience.ro}
$\ $
and
Alessio~Orlandi$^{ab}$\thanks{E-mail: orlandi@bo.infn.it},
\\
\\
{\em $^a$Dipartimento di Fisica e Astronomia, Universit\`a di Bologna}
\\
{\em via Irnerio~46, I-40126 Bologna, Italy}
\\
\\
{\em $^b$I.N.F.N., Sezione di Bologna,}
\\
{\em via B.~Pichat~6/2, I-40127 Bologna, Italy}
\\
\\
{\em $^c$ Institute of Space Science, Atomistilor 409, Magurele, Ilfov, Romania} 
}
\begin{document}
\maketitle
\begin{abstract}
We review some features of BEC models of black holes obtained by means of the
horizon wave function formalism.
We consider the Klein-Gordon equation for a toy graviton field coupled to a static
matter current in a spherically symmetric setup. 
The classical field reproduces the Newtonian potential generated by the matter source,
while the corresponding quantum state is given by a coherent superposition of scalar modes
with continuous occupation number. 
An attractive self-interaction is needed for bound states to form, case in which one finds
that (approximately) one mode is allowed, and the system of $N$ bosons can be
self-confined in a volume of the size of the Schwarzschild radius. 
The horizon wave-function formalism is then used to show that the radius of such a
system corresponds to a proper horizon. The uncertainty in the size of the horizon
is related to the typical
energy of Hawking modes: it decreases with the increasing of the black hole mass
(larger number of gravitons), result in agreement with the semiclassical calculations
and which does not hold for a single very massive particle. 
The spectrum of these systems has two components: a discrete ground state of
energy $m$ (the bosons forming the black hole), and a continuous spectrum with
energy $\omega > m$ (representing the Hawking radiation and modelled with a
Planckian distribution at the expected Hawking temperature).
Assuming the main effect of the internal scatterings is the Hawking radiation,
the $N$-particle state can be collectively described by a single-particle
wave-function given by a superposition of a total ground state with energy
$M = N m$ and a Planckian distribution for $E > M$ at the same
Hawking temperature.
This can be used to compute the partition function and find the usual area law
for the entropy, with a logarithmic correction related with the Hawking component.
The backreaction of modes with $\omega > m$  is also shown to reduce the Hawking flux.
The above corrections suggest that for black holes in this quantum state the evaporation
properly stops for vanishing mass.
\end{abstract}









\section{Introduction}
\label{intro}
\par
\noindent
Unreachable, untestable and directly unobservable, black holes are among the most intriguing and discussed
theoretical predictions of modern physics.
Strong evidence exists that these obscure hollows in the space-time structure are indeed generated when the life
of very massive stars comes to an end.
Oppenheimer and co-workers~\cite{OS} were the pioneers in the field of stellar collapse, and since then
the scientific production of this subject has known an endless growth, exposing many obscure points
in General Relativity, and leaving many questions unanswered (see~\cite{joshi,Bekenstein:2004eh}
and references therein, for example).
Problems get even more harduous if Quantum Mechanics is taken into account when considering the 
formation of black holes.
Both at macroscopic and microscopic scales, one usually turns to Thorne's so-called
{\em hoop conjecture\/}~\cite{Thorne:1972ji}, which states that a black hole is formed whenever a given
mass (or equivalent energy) $M$ is entirely confined inside its corresponding gravitational radius $\Rh$.
We recall that, in the simplest cases, this is just the Schwarzschild radius $\Rh=2\,\gn\,M$ where
$\gn=\lp/\mpl$, and $\lp$ and $\mpl$ are the Planck length and mass, respectively, so that
$\hbar=\lp\,\mpl$~\footnote{We shall of course use units with $c=1$.}.
Initially formulated for macroscopic black holes~\cite{payne}, the hoop conjecture lacks a solid experimental
proof, but from a theoretical perspective it turns out to be solid and reliable on different testing grounds.
In the classical domain of very large systems, the very concept of a background metric and related horizon
structure are well understood (for a review of some problems, see the bibliography in
Ref.~\cite{Senovilla:2007dw}).
Going down to the Planck scale, however, concepts such as the particle's ``size'' and ``position'' become
fuzzy and this makes the hoop conjecture much more difficult to state clearly.
In this range quantum effects are relevant (for a recent discussion, see, e.g., Ref.~\cite{acmo}) and one
should also consider the possible existence of new objects, usually referred to as ``quantum black holes''
(see, e.g., Refs.~\cite{hsu,calmet}).
The idea that black holes may originate in particle collisions is supported by the hoop conjecture itself. 
Two particles (with - say - negligible spatial extension and total angular momentum) can collide with an
impact parameter $b$ shorter than the Schwarzschild radius corresponding to the total center-of-mass
energy $E$ of the system,
\be
b
\leq
2\,\lp\,\frac{E}{\mpl}
\equiv
\Rh(E)
\ .
\label{hoop}
\ee
One then expects that a microscopic black hole is formed, with non-negligible quantum properties.
Because of this, black holes would require a quantum theory of gravity in order to be described and
studied properly. 
Such a (universally accepted) theory does not yet exist and one has to work under (sometimes strong)
approximations.
Quantum Field Theory on curved space-time backgrounds has been developed for this purpose:
this theoretical approach relies on the assumption that quantum gravitational fluctuations are small,
and has produced remarkable results~\cite{hawking}.
\par
A new theoretical model was recently proposed by Dvali and Gomez~\cite{DvaliGomez}, which
allows to describe the quantum properties of black holes from an entirely new perspective:
large black holes are viewed as graviton condensates at a critical point and can be reliably described
on flat space, the curved metric emerging as a collective effect. 
The aim of this review is to present some of the features of this model obtained from the
``horizon wave-function formalism'' (HWF).
In this general approach, a wave-function for the gravitational radius can be associated with any localised
Quantum Mechanical
particle~\cite{Cfuzzy, Casadio:2013aua,qhoop,Casadio:2014twa,Casadio:2015rwa,Casadio:2015sda},
which makes it easy to formulate in a quantitative way a condition for distinguishing black holes from
regular particles.
Among the added values, the formalism naturally leads to an effective Generalised Uncertainty
Principle (GUP)~\cite{scardigli} for the particle's position~\footnote{Similar features are naturally
encoded in non-commutative models of black holes.
For a review, see Ref.~\cite{PNC}.},
and a decay rate for microscopic black holes.
\par
In the next section, we briefly review the corpuscular model of Refs.~\cite{DvaliGomez}
and one of its possible field theoretic implementations in Section~\ref{star}.
(The particular case we consider there will allow us to make some conjectures regarding the
black hole formation from the gravitational collapse of a star.)
In Section~\ref{secHWF}, we review the HWF and how it reproduces a GUP.
This formalism is then applied to specific models of corpuscular black holes and their
Hawking radiation in Section~\ref{secBECbh}, which is the main part of this review,
and collects results from Refs.~\cite{Casadio:2014vja} and~\cite{Casadio:2015bna}.
Finally, some comments and speculations are summarised in Section~\ref{secC}.
\section{Corpuscular model}
\label{CorpMod}
\par\noindent
The simple and intuitive corpuscular model recently introduced by Dvali and Gomez in Refs.~\cite{DvaliGomez},
and widely developed in Refs.~\cite{Casadio:2015bna, Mueck, Foit:2015wqa, Hofmann, Kuhnel}, puts gravitation
under a new light.
The model is based on the assumption that the classical geometry should be viewed as an effective description
of a quantum state with a large graviton occupation number, where gravitons play the role of space-time quanta,
very much like photons are light quanta in a laser beam. 
Unlike photons, which do not interact with each other via quantum electrodynamics, graviton-graviton interaction
is mediated by gravitation itself, whose attractive nature can thus lead to form a ball of superposed gravitational
quanta.
When such a superposition is a ground-state, the gravitational field is effectively a Bose-Einstein condensate (BEC):
Dvali and Gomez conjectured this is precisely what happens inside black holes.
Even when considering a strong gravitational regime, as expected to happen at the verge of black hole formation,
the whole construction can be nicely explained under Newtonian approximation.
The Newtonian potential at a distance $r$ generated by a system of $N$ gravitons, each with effective mass $m$
(so that the total mass $M = N\,m$) is
\be
V_{\rm N}(r)
\simeq
-\frac{\gn\,M}{r}
=
- \frac{\lp \,N\, m}{r\,\mpl}
\ .
\label{Vn}
\ee
This potential can be strong enough to confine the gravitons themselves inside a finite volume where they are all
superposed on each other.
The gravitons effective mass $m$ can be related to their characteristic quantum mechanical size via the
Compton/de~Broglie wavelength $\lambda_m \simeq \frac{\hslash}{m} = \lp\,\frac{\mpl}{m}$.
If one assumes that the interaction is negligible outside the ball of gravitons and constant inside, with average
interaction distance $r=\lambda_{m}$, the potential simplifies to
\be
V_{\rm N}(r) \simeq -\frac{\gn\,M}{\lambda_{m}}\Theta (\lambda_{m}-r) := V_{\rm N}(\lambda_{m})
\ ,
\ee
which results in an average potential energy per graviton
\be
U_m
\simeq
m\,V_{\rm N}(\lambda_m)
:=
-N\,\alpha\,m
\ ,
\label{eq:Udvali}
\ee
where
\be
\alpha = \frac{\lp^{2}}{\lambda_m^{2}} = \frac{m^{2}}{\mpl^{2}} \ , \label{alpha}
\ee
is the gravitational coupling.
\par
One can think of virtually superposing gravitons one by one, thus strengthening their reciprocal attraction,
until they find themselves confined inside a deep enough ``potential well'' from which they cannot escape.
The condition for the gravitons to be ``marginally bound'' is reached when each single graviton has just not
enough energy $E_{K}=m$ to escape the potential well,
\be
E_K + U
\simeq
0
\ .
\label{eq:energy0}
\ee
At this point, one has created a black hole solely out of condensed gravitational quanta.
When this condition is reached, the gravitons are ''maximally packed'', and their number satisfies
\be
N\,\alpha
\simeq
1
\ .
\label{eq:maxP}
\ee
The effective graviton mass correspondingly scales as
\be
m
\simeq
\frac{\mpl}{\sqrt{N}}
\ ,
\ee
while the total mass of the black hole scales like~\footnote{This scaling relation had been already found
in~\cite{Ruffini:1969qy} without fully understanding its role in the case of black hole formation.}
\be
M
=
N\,m
\simeq
\sqrt{N}\,\mpl
\ .
\label{eq:Max}
\ee
Moreover, the horizon's size, namely the Schwarzschild radius
\be
\Rh=2\,\lp\frac{M}{\mpl}
\ ,
\label{RH}
\ee
is spontaneously quantised as commonly expected~\cite{Bekenstein:1997bt}, that is
\be
\Rh
\simeq
\sqrt{N}\, \lp
\ .
\label{eq:areaquantization}
\ee
\par
This simple, purely gravitational black hole can now be shown to emit purely gravitational
Hawking radiation. 
In a first order approximation, reciprocal $2 \to 2$ graviton scatterings inside the condensate will
give rise to a depletion rate 
\be
\Gamma
\sim
\frac{1}{N^{2}}\,N^{2}\,\frac{1}{\sqrt{N}\,\lp}
\ ,
\label{eq:Grate}
\ee
where the factor $N^{-2}$ comes from $\alpha^2$, the second factor is combinatoric
(there are about $N$ gravitons scattering with other $N - 1 \simeq N$ gravitons),
and the last factor comes from the characteristic energy of the process $\Delta E\sim m$.
The amount of gravitons in the condensate will then decrease according
to~\cite{DvaliGomez}
\be
\dot N
\simeq
- \Gamma
\simeq
-\frac{1}{\sqrt{N}\, \lp} + {O}(N^{-1})
\ .
\label{eq:depleted-N}
\ee
As explained in Refs.~\cite{DvaliGomez}, this emission of gravitons reproduces
the purely gravitational part of the Hawking radiation and contributes to the shrinking
of the black hole according to the standard results
\be
\dot M
\simeq
\mpl\,\frac{\dot N}{\sqrt{N}}
\sim
-\frac{\mpl}{N\, \lp}
\sim
-\frac{\mpl^3}{\lp\,M^2}
\ .
\label{dotM}
\ee
From this flux one can then read off the ``effective'' Hawking temperature 
\be
T_{\rm H}
\simeq
\frac{\mpl^2}{8\,\pi\,M}
\sim
m
\sim
\frac{\mpl}{\sqrt{N}}
\ ,
\label{T_H}
\ee
where the last expression is precisely the approximate value we shall
use throughout.
\par
Let us remark once more that this BEC model only contains gravity and
the Hawking radiation we consider here is therefore just the graviton contribution.
In a more refined model, one would also like to include all possible
matter content (presumably originating from the astrophysical object that
collapsed to form the black hole) and hopefully recover all types of
radiation~\footnote{For some preliminary results regarding the role of baryons,
see Ref.~\cite{kuhnelBaryons}.}.
\section{Scalar toy-gravitons coupled to a source}
\label{star}
\par\noindent
The model presented in the previous section is very simple and leads to very ``reasonable''
properties, but in its original form lacks some features that might make it even more appealing.
For example, the connection with the usual geometrical picture of General Relativity is not immediate,
and the horizon ``emerges'' from a classical mechanical condition (the binding condition
of Eq.~\eqref{eq:energy0}), rather than from relativistic considerations (as it happens in the case of the
hoop conjecture).
\par
Trying to understand the corpuscular theory of Dvali and Gomez, using different tools can be of some help.
In Refs.~\cite{Casadio:2014vja}, Quantum Field Theory was proposed in order to model a self-sustained
graviton system.
To simplify the description, the authors consider scalar toy-gravitons instead of regular gravitons,
which allows to employ the Klein-Gordon equation for a real and massless scalar field $\phi$ coupled
to a real scalar current $J$ in Minkowski space-time,
\be
\Box\phi(x)
=
q\,J(x)
\ ,
\label{eq:EOM}
\ee
where $\Box=\eta^{\mu\nu}\,\partial_\mu\,\partial_\nu$ and $q$ is
a dimensionless coupling.
One then also assumes that the current is time-independent,
$\partial_0J=0$. 
In momentum space, with $k^\mu=(k^0,\mathbf{k})$, this leads to
\be
k^0\,\tilde{J}(k^\mu)
=
0
\ee
which is solved by the distribution
\be
\tilde{J}(k^\mu)
=
2\,\pi\,\delta(k^0)\,\tilde{J}(\mathbf{k})
\ ,
\ee
where $\tilde{J}^*(\mathbf{k})=\tilde{J}(-\mathbf{k})$.
For the spatial part, exact spherical symmetry is assumed,
so that the analysis is restricted to functions of the kind $f(\mathbf{x})=f(r)$,
with $r=|\mathbf{x}|$.
Classical spherically symmetric solutions of Eq.~\eqref{eq:EOM} can
be formally written as
\be
\phi_{\rm c}(r)
=
q\,\Box^{-1}
J(r)
\ ,
\label{eq:EOMmom}
\ee
and they can be found more easily in momentum space.
The latter is defined by the integral trasnformation
\be
\tilde{f}(k)
=
4\,\pi\int_0^{+\infty}{dr\,r^2\,j_0(kr)\,f(r)}
\ ,
\ee
where  
\be
j_0(kr)
=
\frac{\sin(kr)}{k\,r}
\ ,
\ee
is a spherical Bessel function of the first kind and $k=|\mathbf{k}|$.
This gives the solution 
\be
\tilde{\phi}_{\rm c}(k)
=
-q\,\frac{\tilde{J}(k)}{k^2}
\ .
\label{eqkc}
\ee
\par
For example, for a current with Gaussian profile,
\be
J(r)
=
\frac{e^{-r^2/(2\,\sigma^2)}}{(2\,\pi\,\sigma^2)^{3/2}}
\ ,
\label{eq:Gauss}
\ee
one finds
\be
\tilde{J}(k)
=
e^{-k^2\sigma^2/2}
\ee
and the corresponding classical scalar field is given by
\be
\phi_{\rm c}(r)
&\!\!=\!\!&
-\frac{q}{2\,\pi^2}\int_0^{+\infty}{\d k\,
j_0(kr)\,e^{-k^2\sigma^2/2}}
\nonumber
\\
&\!\!=\!\!&
-\frac{q}{4\,\pi\,r}\,
\mathrm{erf}{\left(\frac{r}{\sqrt{2}\,\sigma}\right)}
\ ,
\ee
where $\mathrm{erf}$ is the error function.
At large distances from the source $J$, when $r\gg\sigma$,
the field $\phi$ reproduces the classical Newtonian potential~\eqref{Vn},~i.e. 
\be
V_{\rm N}
=
\frac{4\,\pi}{q}\,\gn\,M\,\phi_{\rm c}
\simeq
-\frac{\gn\,M}{r}
\ 
\label{V_N}
\ee
\par
In the quantum theory, the classical configurations~\eqref{eq:EOMmom}
are replaced by coherent states.
To prove this statement, one can start with the normal-ordered quantum Hamiltonian density
in momentum space,
\be
\hat{\mathcal{H}}
=
k\,\hat a'^\dagger_k\,\hat a'_k+\tilde{\mathcal{H}}_g
\ ,
\label{eq:hamquant}
\ee
where $\tilde{\mathcal{H}}_g$ is the ground state energy density, 
\be
\tilde{\mathcal{H}}_g
=
-q^2\,\frac{|\tilde{J}(k)|^2}{2\,k^2}
\ ,
\ee
and the standard ladder operators are shifted according to
\be
\hat a'_k
=
\hat a_k+q\,\frac{\tilde{J}(k)}{\sqrt{2\,k^3}}
\ .
\label{eq:laddshift}
\ee
The source-dependent ground state $\Ket{g}$ is annihilated by the
shifted annihilation operator,
\be
\hat a'_k
\Ket{g}
=
0
\ ,
\ee
and is a coherent state in terms of the standard field vacuum,
\be
\hat a_k
\Ket{g}
=
-q\,\frac{\tilde{J}(k)}{\sqrt{2\,k^3}}
\Ket{g}
=
g(k)
\Ket{g}
\ ,
\ee
with $g=g(k)$ being an eigenvalue of the shifted annihilation operator.
This results in
\be
\Ket{g}
=
e^{-N/2}\,
{\rm {exp}}\left\{\int{\frac{k^2\,\d k}{2\,\pi^2}\,g(k)\,\hat a^\dagger_k}\right\}
\Ket{0}
\ ,
\label{ketg}
\ee
with $N$ representing the expectation value of the number of quanta
in the coherent state,
\be
N
&\!\!=\!\!&
\int{\frac{k^2\,\d k}{2\,\pi^2}\,
\Bra{g}\hat a^\dagger_k\,\hat a_k\Ket{g}}
\nonumber
\\
&\!\!=\!\!&
\frac{q^2}{(2\,\pi)^2}
\int \frac{\d k}{k}\,
|\tilde{J}(k)|^2
\ ,
\label{expN}
\ee
from which the occupation number is found to be
\be
n_k
=
\left(\frac{q}{2\,\pi}\right)^2
\frac{|\tilde{J}(k)|^2}{k}
\ .
\ee
It is now straightforward to verify that the expectation value of the field in the
state $\Ket{g}$ coincides with its classical value,
\be
\Bra{g}\hat {\phi}_k\Ket{g}
&\!\!=\!\!&
\frac{1}{\sqrt{2\,k}}
\Bra{g}
\left(\hat a_k+\hat a^\dagger_{-k}\right)
\Ket{g}
\nonumber
\\
&\!\!=\!\!&
\frac{1}{\sqrt{2\,k}}
\Bra{g}
\left(\hat a'_k+\hat a'^\dagger_{-k}\right)
\Ket{g}
-q\,\frac{\tilde{J}(k)}{k^2}
\nonumber
\\
&\!\!=\!\!&
\tilde \phi_{\rm c}(k)
\ ,
\label{expPc}
\ee
thus $\ket{g}$ is a realisation of the Ehrenfest theorem.
\par
It is important to note that Eq.~\eqref{expN} presents an~UV divergence if the source
has infinitely thin support, and an IR divergence if the source contains modes of
vanishing momenta (which would only be physically consistent with an eternal source).
Because to this, the state $\Ket{g}$ and the number $N$ are not mathematically
well-defined in general.
Anyway, the UV divergence can be cured, for example, by using a Gaussian
distribution like the one in Eq.~\eqref{eq:Gauss}, while the IR divergence can
be eliminated if the scalar field is massive or the system is enclosed within a finite
volume (so that allowed modes are also quantised). 
\subsection{Black holes as self-sustained quantum states}
\label{BH}
\label{2.1}
\par\noindent
One can now analyse a ``star'', made of ordinary matter whose density is distributed according to
\be
\rho=M\,J
\ ,
\label{rhoM}
\ee
where $M$ is the total (proper) energy of the star.
Its Newtonian potential energy is given by $U_{M}=M\,V_{\rm N}$,
so that $\phi_{\rm c}$ is accordingly determined by Eq.~\eqref{V_N} and the quantum state
of $\hat \phi$ by Eq.~\eqref{expPc}.
\par
During the formation of a black hole, the gravitons are expected to dominate
the dynamics over the matter source~\cite{DvaliGomez}.
Then, one can assume the matter contribution is negligible,
and the source $J$ in the r.h.s.~of Eq.~\eqref{eq:EOM}
is thus provided by the gravitons themselves.
This source, consisting of gravitons, is roughly confined in a finite spherical volume
$\mathcal{V}=4\,\pi\,R^3/3$ (this is a crucial feature for the ``classicalization'' of
gravity~\cite{dvaliCL} and requires an attractive self-interaction for the scalar field to admit bound states).
The energy density~\eqref{rhoM} must then be equal to the
average energy density inside the volume $\mathcal{V}$, which in turn is given by the
average potential energy of each graviton in $\mathcal{V}$, times the number of gravitons:
$N\,U_m/\mathcal{V}$.~\footnote{Each graviton interacts with the other $N-1$, so that the energy of each
graviton is proportional to $(N-1)\,U_m$, but one can safely approximate $N-1\simeq N$, given that $N$
is considered large.}
This assumption is qualitatively the same as the marginally bound condition~\eqref{eq:energy0}
with $N\,E_{\rm K}\sim J$.
After some simple substitutions one finds
\be
J
\simeq
-\frac{3\,N\,\gn\,m}{q\,R^3}\,\phi_{\rm c}
\ ,
\label{Jg}
\ee
inside the volume $\mathcal{V}$, where $m$ is the energy of each of the $N$ scalar
gravitons.
Using this condition into Eq.~\eqref{eqkc}, one finds
\be
\frac{3\,N\,\gn\,m}{R^3\,k^2}
=
\frac{3\,\Rh}{2\,R^3\,k^2}
\simeq
1
\ ,
\ee
where
\be
\Rh=2\,\gn\,M
\label{Rh}
\ee
is the classical Schwarzschild radius of the object.
One can infer that a self-sustained system of gravitons will contain only the modes
with momentum numbers $k=k_{\rm c}$ such that
\be
R\,k_{\rm c}
\simeq 
\sqrt{\frac{\Rh}{R}}
\ ,
\label{kc}
\ee
where numerical coefficients of order one were dropped in line with
the qualitative nature of the analysis.
\par
This clearly does not happen in the Newtonian case, where the potential generated by an ordinary
matter source would allow any momentum numbers.
Ideally, this means that, if the gravitons represent the main self-gravitating source,
the quantum state of the system must be given in terms of just one mode
$\phi_{k_{\rm c}}$.
A coherent state of the form in Eq.~\eqref{ketg} requires a distribution of different momenta
and cannot thus be built this way, i.e.~by means of a strictly confined source.
Furthermore, the relation~\eqref{expN} between $N$ and the source momenta
does not apply here.
Instead, for very large $N$, all scalars are in the state $\ket{k_{\rm c}}$
and form a BEC.
\par
Consider now that, for an ordinary star, the typical size is much greater than its Schwarzschild
radius ($R\gg \Rh$) and therefore $k_{\rm c}\ll R^{-1}$.
The corresponding de~Broglie length $\lambda_{\rm c}\simeq k^{-1}_{\rm c}\gg R$, 
which conflicts with the assumption that the field represents a gravitating source
only within a region of size $R$.
However, in the ``black hole limit'' $R\sim\Rh$, and recalling that $m=\hbar\,k$, one obtains 
\be
1
\simeq
\gn\,M\,k_{\rm c}
=
N\,\frac{m^2}{\mpl^2}
\ ,
\label{1=N}
\ee
which leads to the two scaling relations~\eqref{eq:Max}, namely
$m=\hbar\,k_{\rm c}\simeq \mpl/\sqrt{N}$ 
and a consistent de~Broglie length $\lambda_m\simeq\lambda_{\rm c}\simeq \Rh$.
Therefore, an ideal system of self-sustained (toy) gravitons must be a BEC with a size
that suggests it is a black hole.
All that remains to be proven is the existence of a horizon, or at least a trapping surface,
in the given space-time.
\par
The safest way to find trapping surfaces would require a general-relativistic solution
for the self-gravitating BEC with given density and equation of state.
Many authors faced this problem in the past decades.
Self-gravitating boson stars in General Relativity have been studied for example
in Ref.~\cite{Ruffini:1969qy}, but only numerical solutions have been found, mostly
generated by a Gaussian source~\cite{Colpi:1986ye,Chavanis:2011cz}.
In the specific case presented here, Quantum Mechanics crosses the way of General Relativity
because the BEC black hole can be regarded to as a ``giant soliton'', and
microscopic black holes are extremely dense, thus producing a (relatively) strong space-time curvature
at very small scales.
In these regimes it is uncertain whether a semiclassical approach is still valid,
since the quantum fluctuations become relevant with respect to the surrounding
space-time geometry.
The HWF~\cite{Cfuzzy, Casadio:2013aua, qhoop, Casadio:2014twa, Casadio:2015rwa, Casadio:2015sda}
was precisely proposed in order to define the gravitational radius of any quantum system,
and should therefore be very useful for investigating this issue.
\par
To avoid misunderstandings, one needs to make some clarifications about the toy model
presented above (where the field $\phi$ is totally confined inside a sphere of radius $\Rh$).
The scaling relation~\eqref{1=N} does not require the scalar field $\phi$ to vanish (or be negligible)
outside the region of radius $\Rh$.
Since the scalar field also provides the Newtonian potential (see Eq.~\eqref{V_N}),
the vanishing of the field outside $\Rh$ would imply there is no Newtonian potential
outside the BEC, and this would conflict with the idea that the BEC is a gravitational source.
To recover the classical Newtonian potential $V_{\rm N}\sim\phi$ outside the source,
it is enough to relax the condition~\eqref{Jg} for $r\gtrsim \Rh$ (where $J\simeq 0$),
and properly match the (expectation) value of $\hat\phi$ with the Newtonian $\phi_{\rm c}$
from Eq.~\eqref{V_N} at $r\gtrsim \Rh$.
One then expects that for $r\gg\Rh$ and $N\gg 1$, the classical description is recovered
and that the total mass $M$ of the black hole becomes the only relevant quantity.
A hint of this can be found in the classical analysis of the outer ($r\gg\sigma\sim\Rh$)
Newtonian scalar potential and its quantum counter-part,
but also in the alternative description of gravitational scattering.
Geodesic motion can be reproduced in the post-Newtonian expansion of the Schwarzschild metric
by tree-level Feynman diagrams with graviton exchanges between a test probe
and a (classical) large source~\cite{duff}~\footnote{For a similarly non-geometric
derivation of the action of Einstein gravity, see Ref.~\cite{deser}.}.
For $r\gg\Rh$, the source in this calculation is described by its total mass $M$,
and quantum effects should then be suppressed by factors of $1/N$~\cite{DvaliGomez}.
\section{Horizon wave-function formalism}
\label{secHWF}
\par\noindent
So far, space-time is assumed to be flat and there are no general relativistic
effects, more exactly there is no hint of a non-trivial causal structure around
the BEC.
From an observer's perspective, nothing happens when crossing the surface
of the BEC sphere, so is this system a black hole in the usual sense?
One can show that such a system of $N\gg 1$ gravitons is a black hole
in the usual sense, by identifying its actual gravitational radius and then arguing 
it represents a trapping surface.
\par
A fairly recent formalism which is very suitable for such a task is the HWF introduced
in Ref.~\cite{Cfuzzy} and further developed in
Refs.~\cite{Casadio:2013aua, qhoop, Casadio:2014twa, Casadio:2015rwa, Casadio:2015sda}.
This tool applies to the quantum mechanical state $\psi_{\rm S}$ of any
system {\em localised in space\/} and {\em at rest\/} in the chosen reference frame.
The HWF formalism was first formulated for a single particle, but extended to a system
of large-$N$ particles to correctly describe a macroscopic black hole.
One starts by defining suitable Hamiltonian eigenmodes,
\be
\hat H\,\ket{\psi_E}=E\,\ket{\psi_E}
\ , 
\ee
where $H$ can be specified depending on the model we wish to consider.
The state $\psi_{\rm S}$ is initially decomposed in energy eigenstates
\be
\ket{\psi_{\rm S}}
=
\sum_E\,C(E)\,\ket{\psi_E}
\ .
\label{CE}
\ee
For the simple example of an electrically neutral object with spherical symmetry,
by inverting the expression of the Schwarzschild radius,
\be
\rh
=
2\,\gn\,E = 2\, \frac{\lp}{\mpl}E
\ ,
\label{rh}
\ee
one obtains $E$ as a function of $\rh$.
This expression is then used to define the {\em horizon wave-function\/} as
\be
\psi_{\rm H}(\rh)
\propto
C(E) = C\left(\mpl\,{\rh}/{2\,\lp}\right)
\ ,
\ee
whose normalisation is determined by the inner product
\be
\pro{\psi_{\rm H}}{\phi_{\rm H}}
=
4\,\pi\,\int_0^\infty
\psi_{\rm H}^*(\rh)\,\phi_{\rm H}(\rh)\,\rh^2\,\d \rh
\ .
\label{Hpro}
\ee
Starting from the HWF one then goes to define the probability density of finding a gravitational radius $r=\rh$
corresponding to the given quantum state $\psi_{\rm S}$ as
\be
{\mathcal P}_{\rm H}(\rh)=4\,\pi\,\rh^2\,|\psi_{\rm H}(\rh)|^2
\ .
\ee
It is now clear that the HWF takes the classical concept of the Schwarzschild radius into the quantum domain. 
The gravitational radius turns out to be necessarily ``fuzzy'', since it is related to a quantity
(the energy of the particle) that is naturally uncertain.
\par
The next step is to calculate the probability density for the particle to be inside its own gravitational
radius $r=\rh$,
\be
{\mathcal P}_<(r<\rh)
=
P_{\rm S}(r<\rh)\,{\mathcal P}_{\rm H}(\rh)
\ ,
\label{PrlessH}
\ee
where
\be
P_{\rm S}(r<\rh)
=
4\,\pi\,\int_0^{\rh}
|\psi_{\rm S}(r)|^2\,r^2\,\d r
\ee
is the probability that the system is inside a sphere of radius $r=\rh$, and $P_{\rm H}(r_{\rm H})$ acts
as a statistical weight for the given gravitational radius.
Now, the probability for the particle described by the wave-function $\psi_{\rm S}$ to be contained
inside its gravitational radius is calculated by integrating~\eqref{PrlessH} over all values of $r_{\rm H}$,
\be
P_{\rm BH}
=
\int_0^\infty {\mathcal P}_<(r<\rh)\,\d \rh
\ .
\label{PBH}
\ee
Therefore, $P_{\rm BH}$ is the probability for a trapping surface (i.e.: a horizon) to exist,
and for the object to be viewed as a black hole.
\par
One of the simplest cases, discussed analytically in Refs.~\cite{Cfuzzy,Casadio:2013aua},
is that of a particle described by the spherically symmetric Gaussian wave-function
\be
\psi_{\rm S}(r)
=
\frac{e^{-\frac{r^2}{2\,\ell^2}}}{\ell^{3/2}\,\pi^{3/4}}
\ ,
\label{psis}
\ee
where the width $\ell$ is assumed to be the minimum 
compatible with the Heisenberg uncertainty principle
$
\ell
=
\lambda_m
\simeq
\lp\,\frac{\mpl}{m}
$ and $\lambda_m$ is the Compton length of the particle of rest mass $m$. 
Moreover, we shall assume the flat space dispersion relation $E^2=p^2+m^2$,
where $p$ is the radial momentum.
This assumption is in line with the general relativistic theory of spherically symmetric
systems, for which the total energy $E$ (known as Misner-Sharp mass) is in fact 
defined as if the space were flat (for more details about this rather subtle point,
see Refs.~\cite{Cfuzzy,Casadio:2013aua,Casadio:2014twa}). 

\begin{figure}[h!tb]
\centering
\raisebox{5.7cm}{${\mathcal P}_<$}
\includegraphics[width=0.6\textwidth]{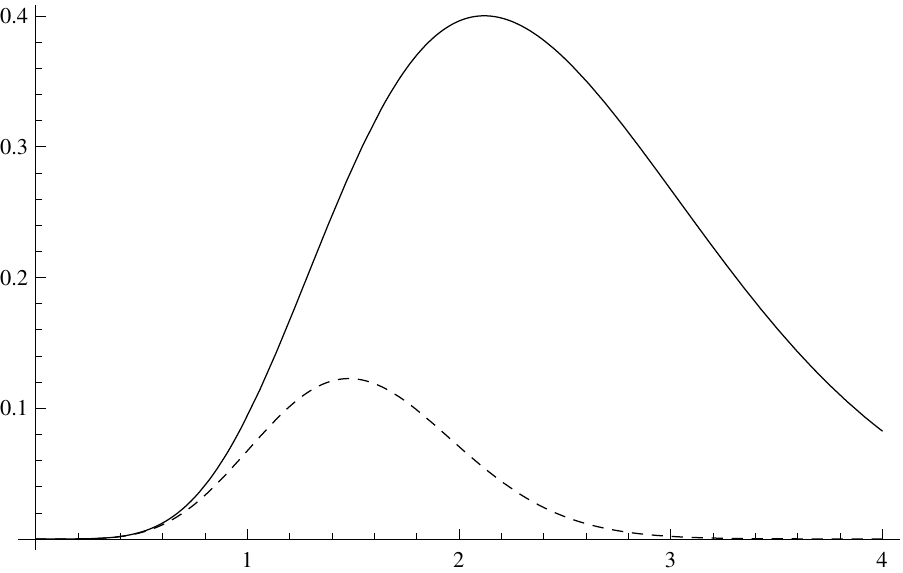}
\\
\hspace{8cm}$\rh/\lp$
\caption{Probability density ${\mathcal P}_<$ in Eq.~\eqref{PrlessH} that particle is inside
its horizon of radius $R=\rh$, for $\ell=\lp$ (solid line) and for $\ell=2\,\lp$ (dashed line).
\label{prob<}}
\end{figure}
\begin{figure}[h!tb]
\centering
\raisebox{5.7cm}{$P_{\rm BH}$}
\includegraphics[width=0.6\textwidth]{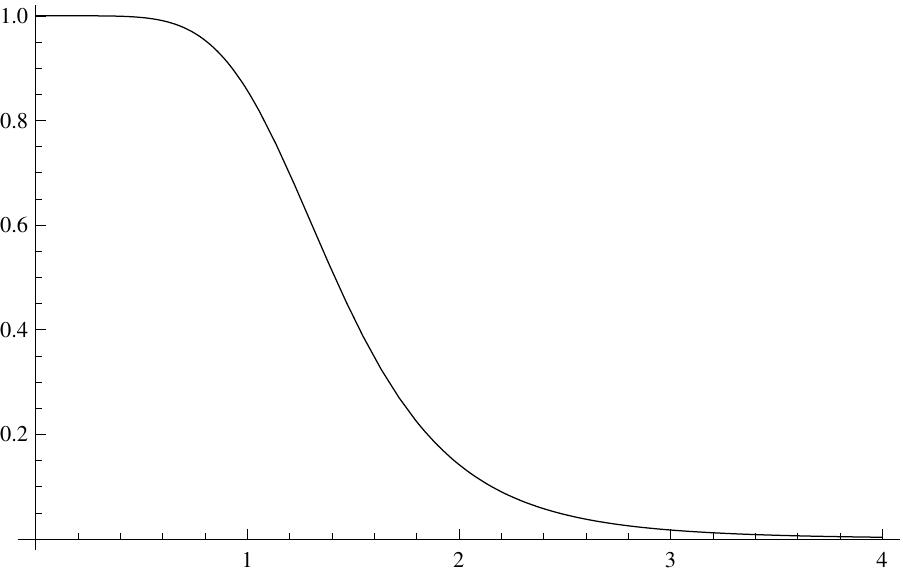}
\\
\hspace{8cm}$\ell/\lp$, $\mpl/m$
\caption{Probability $P_{\rm BH}$ in Eq.~\eqref{PBH} that particle of width $\ell\sim m^{-1}$ is
a black hole.
\label{prob}}
\end{figure}
The probability density ${\mathcal P}_<$ for this particular case is represented graphically in
Fig.~\ref{prob<} for two values of the Gaussian width: $\ell=\lp$ and $\ell=2\,\lp$
(corresponding to $m=\mpl$ and $m=0.5\, \mpl$).
As expected, larger values of the particle mass (narrower spread of the wave-packet) result
in a larger probability density. 
By integrating the probability density, one obtains the probability $P_{\rm BH}$ for the particle
to be a black hole, as stated in \eqref{PBH}. 
Fig. \ref{prob} shows $P_{\rm BH}$ as a function of $\ell$ (in units of $\lp$). 
Gaussian wave-packets with widths below the Planck length are most likely black holes,
but this probability decreases and it becomes negligible for widths of about $3\,\lp$.
The smooth transition from $P_{\rm BH} = 1$ to $P_{\rm BH} = 0$ in a range of mass values
between $\mpl$ and $1/3\, \mpl$ is a feature of this quantum mechanical treatment of the horizon. 
\subsection{Effective GUP}
\label{GUP}
\par\noindent

\begin{figure}[t]
\centering
\raisebox{5.7cm}{$\frac{\Delta r}{\lp}$}
\includegraphics[width=0.6\textwidth]{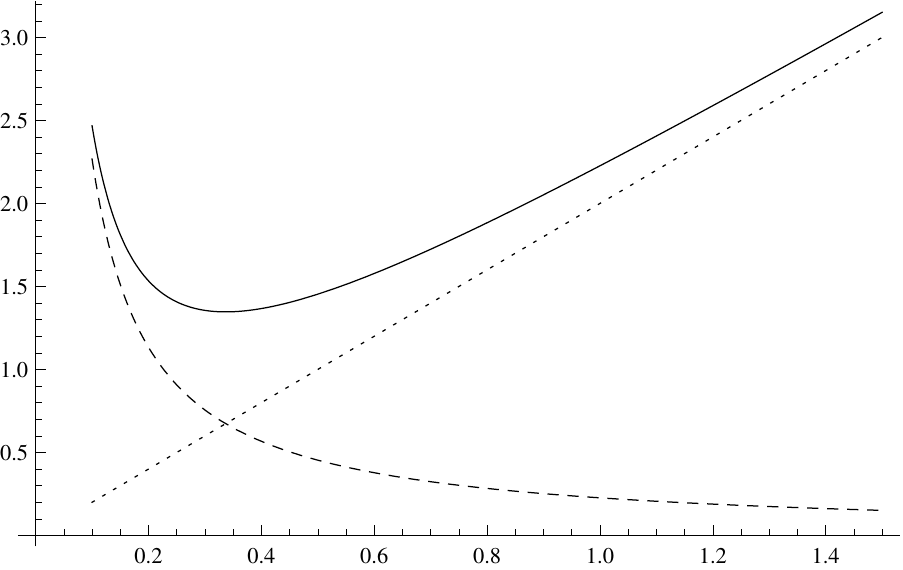}
\\
\hspace{8cm}$\Delta p/\mpl$
\caption{Uncertainty relation~\eqref{effGUP} (solid line) as a combination
of the quantum mechanical uncertainty (dashed line) and the uncertainty
in horizon radius (dotted line).
\label{pGUP}}
\end{figure}

It was mentioned in Section~\ref{intro} that the HWF naturally leads to a GUP. 
For wave-packets near the Planck scale there are two sources of uncertainty:
the standard quantum mechanical uncertainty, and the uncertainty in the size
of the horizon radius (the complete details can be found in Ref.~\cite{Casadio:2013aua}).
By linearly combining the two, we find
\be
\Delta r
&\!\!\equiv\!\!&
\sqrt{\expec{\Delta r^2}}
+
\gamma\,
\sqrt{\expec{\Delta \rh^2}}
\nonumber
\\
&\!\!=\!\!&
\left(\frac{3\,\pi-8}{2\,\pi}\right)
\lp\,\frac{\mpl}{\Delta p}
+
2\,\gamma\,\lp\,\frac{\Delta p}{\mpl}
\ ,
\label{effGUP}
\ee
where $\gamma$ is a coefficient of order one.
The result is plotted (for $\gamma=1$) in Fig.~\ref{pGUP}, where it is also compared
to the usual Heisenberg uncertainty in the position.
\par
The uncertainty in the size the horizon for trans-Planckian wave-packets 
(with masses $M\gg\mpl$) turns out to be
\be
\Delta\rh
\sim
\expec{\hat r_{\rm H}}
\sim
\Rh
\ .
\label{Mmp}
\ee
Such large quantum fluctuations, with an amplitude comparable to the size of the classical
gravitational radius, can hardly be considered appropriate for any classical system.
Therefore, one comes to the striking conclusion that this type of source cannot
represent large, semiclassical black holes, for which the relative uncertainty is expected
to decrease to zero very fast for increasing mass.
\section{BEC black hole horizons}
\label{secBECbh}
\par\noindent

When analysing the case of BEC black holes, one has to remember that these are
composed of very large numbers of particles (the toy-gravitons) of very small effective
mass $m\ll \mpl$ (thus very large de~Broglie length, $\lambda_m\gg\lp$), as opposed
to the case of single massive particles studied in
Refs.~\cite{Casadio:2013aua, qhoop, Casadio:2014twa, Casadio:2015rwa, Casadio:2015sda}
and briefly reviewed in Section~\ref{secHWF}. 
According to Ref.~\cite{Casadio:2013aua}, they cannot individually form (light) black holes.
A generalisation of the formalism to a system of $N$ such components is possible and
will allow one to show that the total energy $E=M$ is sufficient to create a proper horizon.
\par
In order to set the notation, consider a system of $N$ scalar particles (our ``toy gravitons''),
$i=1,\ldots,N$, whose dynamics is determined by a Hamiltonian $H_i$.
If the particles are marginally bound according to Eq.~\eqref{eq:energy0}, the single-particle
Hilbert space can be assumed to contain the discrete ground state $\ket{m}$, defined by
\be
\hat H_i\Ket{m}
=
m\Ket{m}
\ ,
\ee
and a gapless continuous spectrum of energy eigenstates $\Ket{\omega_i}$,
such that
\be
\hat H_i\Ket{\omega_i}
=
\omega_i\Ket{\omega_i}
\ ,
\ee
with $\omega_i>m$.
The continuous spectrum reproduces the particles that escape the BEC.
Each particle is then assumed to be in a state given by a superposition
of $\ket{m}$ and the continuous spectrum, namely 
\be
\ket{\Psi_{\rm S}^{(i)}}
=
\frac{\ket{m}+\gamma_1\ket{\psi^{(i)}_{\rm S}}}
{\sqrt{1+\gamma_1^2}}
\ ,
\ee
where $\gamma_1 \in \mathbb{R}_+$ is a dimensionless parameter that weights
the relative probability amplitude for each ``toy graviton'' to be in the continuum
rather than ground state.
The total wave-function will be given by the symmetrised product of $N$ such states,
\be
\ket{\Psi_N}
\simeq
\frac{1}{N!}\,
\sum_{\{\sigma_i\}}^N
\left[
\bigotimes_{i=1}^N
\,
\ket{\Psi_{\rm S}^{(i)}}
\right]
\ ,
\label{Nstate}
\ee
where the sum is over all the permutations $\{\sigma_i\}$ of the $N$ excitations.
Since the interaction is included into terms proportional to powers of $\gamma_1$,
the spectral decomposition of this $N$-particle state can be obtained by defining the
total Hamiltonian simply as the sum of $N$ single-particle Hamiltonians, 
\be
\hat H
=
\bigoplus_{i=1}^N
\hat H_i
\ .
\ee
The corresponding eigenvector for the discrete ground state with $M=N\,m$ is given by
\be
\hat H\Ket{M}
=
M\Ket{M}
\ ,
\ee
and the eigenvectors for the continuum by
\be
\hat H\Ket{E}
=
E\Ket{E}
\ .
\ee
The spectral coefficients are computed by projecting $\ket{\Psi_{\rm N}}$ on 
these eigenvectors.
\subsection{Black holes with no hair}
\par
\label{idealBH}
\par\noindent

\begin{figure}[t]
\centering
\includegraphics[width=14cm]{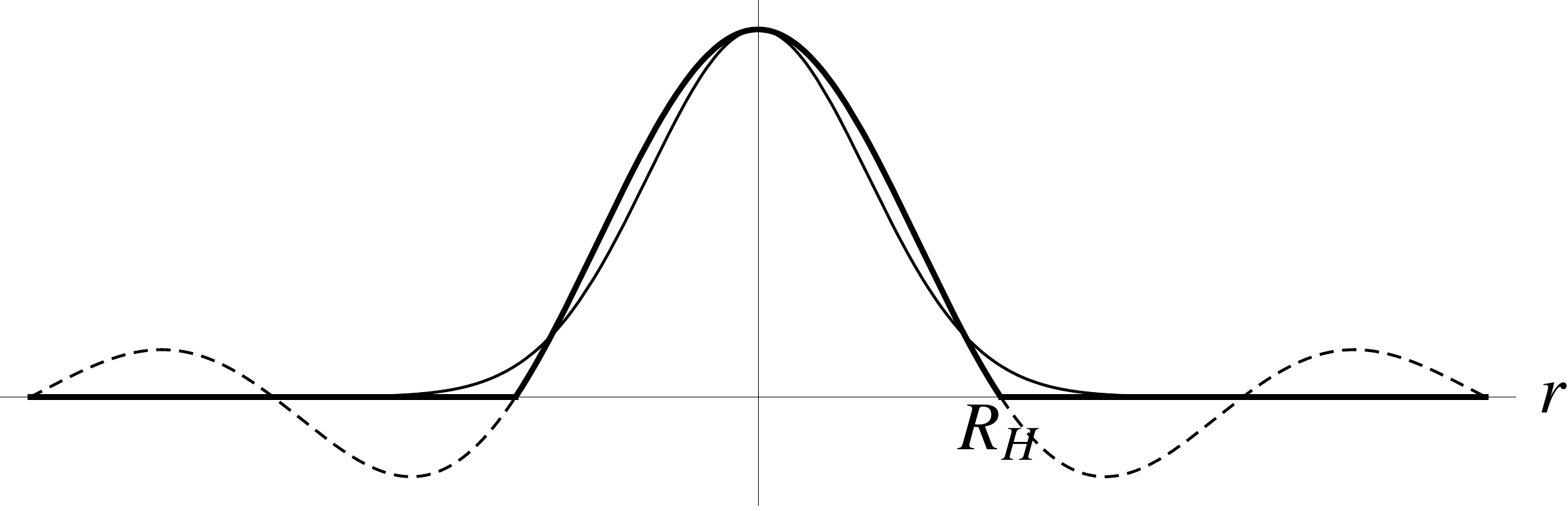}
\caption{Scalar field mode of momentum number $k_{\rm c}$:
ideal approximation in Eq.~\eqref{psiS0} (thick solid line) compared
to exact $j_0(k_{\rm c} r)$ (dashed line).
The thin solid line represents a Gaussian distribution of the kind considered
in Ref.~\cite{Casadio:2013hja}.
\label{ideal}}
\end{figure}

The highly idealised case in Eq.~\eqref{kc} admits precisely one mode, given by 
\be
k_{\rm c}
=
\frac{\pi}{\Rh}
=
\frac{\pi}{2\,\sqrt{N}\,\lp}
\ ,
\ee
so that, on the surface of the black hole, $\phi_{k_{\rm c}}(\Rh)\simeq j_0(k_{\rm c}\,\Rh)=0$,
and the scalar field vanishes outside of $r=\Rh$,
\be
\psi_{\rm S}(r_i)
=
\pro{r_i}{k_{\rm c}}
=
\left\{
\begin{array}{ll}
\mathcal{N}_{\rm c}\,j_0(k_{\rm c} r_i)
&
{\rm for}\ 
r<\Rh
\\
\\
0
&
{\rm for}\ 
r>\Rh
\ ,
\end{array}
\right.
\label{psiS0}
\ee
where $\mathcal{N}_{\rm c}=\sqrt{\pi/2\,\Rh^3}$ is a normalisation factor such that
\be
4\,\pi\,\mathcal{N}_{\rm c}^2
\int_0^{\Rh}
|j_0(k_{\rm c} r)|^2
\,r^2\,\d r
=
1
\ .
\ee
This approximate mode is plotted in Fig.~\ref{ideal}, and compared with a Gaussian distribution
of the kind considered in Ref.~\cite{Casadio:2013hja}. 
\par
It was noted before the end of Section~\ref{2.1} that a scalar field
which vanishes everywhere outside $r=\Rh$ is inconsistent with the existence
of an outer Newtonian potential, but one can still investigate the case for the sake
of having a complete picture.
A system of $N$ such modes will be described by a wave-function which is 
the (totally symmetrised) product of $N\sim M^2$ equal modes and $\gamma_1=0$,
that is
\be
\Psi_{\rm N}(r_1,\ldots,r_N)
=
\frac{\mathcal{N}_{\rm c}^N}{N!}\,
\sum_{\{\sigma_i\}}^N\,
\prod_{i=1}^N
\,
j_0(k_{\rm c} r_i) 
\ .
\ee
This is clearly an eigenstate of the total Hamiltonian
\be
\hat H\,\ket{\Psi_{\rm N}}
=
N\,\hbar\,k_{\rm c}\,\ket{\Psi_{\rm N}}
=
M\,\ket{\Psi_{\rm N}}
\ ,
\ee
and there exists only one non-vanishing coefficient in the spectral decomposition:
$C(E)=1$, for $E=N\,\hbar\,k_{\rm c}=M$, corresponding to a probability
density for finding the horizon size between $\rh$ and $\rh+\d\rh$  
\be
{\mathcal P}_{\rm H}(\rh)
&\!\!=\!\!&
4\,\pi\,
\rh^2\,
\left|\psi_{\rm H}(\rh)\right|^2\,
\nonumber
\\
&\!\!=\!\!&
\delta(\rh-\Rh)
\ .
\ee
This result, and the fact that all the excitations in the mode $k_{\rm c}$
are confined within the radius $\Rh\simeq\lambda_{\rm c}$, leads to the conclusion
that the system is a black hole,
\be
P_{\rm BH}
&\!\!\simeq\!\!&
4\,\pi\,\mathcal{N}_{\rm c}^2
\int_0^\infty
\d\rh\,
\delta(\rh-\Rh)
\int_0^{\rh}
|j_0(k_{\rm c}\,r)|^2
\,r^2\,\d r
\nonumber
\\
&\!\!=\!\!&
P_{\rm S}(r<\Rh)
=
1
\ .
\ee
with the horizon located at its classical radius,
\be
\expec{\hat r_{\rm H}}
\equiv
\bra{\psi_{\rm H}} \hat r_{\rm H}\ket{\psi_{\rm H}}
=\Rh
\ ,
\label{expRh}
\ee
and with absolutely negligible uncertainty
\be
\Delta\rh^2
\equiv
\bra{\psi_{\rm H}}\left(\hat r_{\rm H}^2-\Rh^2\right)\ket{\psi_{\rm H}}\simeq 0
\ .
\ee
The sharply vanishing uncertainty in $\expec{\hat r_{\rm H}}$ is an unphysical result,
which is a consequence of considering the macroscopic black hole as a pure quantum
mechanical state built by superposing many wave-functions of the type~\eqref{psiS0}.
Also, a zero field at $r>\Rh$ cannot reproduce the Newtonian potential outside the black hole,
which clearly contradicts observations.
\par
For a more realistic macroscopic black hole (with $N\gg 1$), one therefore needs to consider
the existence of more modes besides the ones with $k=k_{\rm c}$.
In this case, the modes with $k= k_{\rm c}$ form a discrete spectrum (which comes
from $k_{\rm c}$ being the minimum allowed momentum, in agreement with the idea
of a BEC of gravitons), and must be treated separately.
If modes with $k> k_{\rm c}$ exist, these would not be (marginally) trapped and
could ``leak out'', thus representing a simple modelisation of the Hawking flux.
They will form a continuous spectrum, which will lead to fuzziness in the horizon's location.
The precise form of this part of the spectrum is however an open issue, and we will
review several possibilities in the next sections.
\subsection{Black hole with gaussian excited spectrum}
\label{BHhair}
\begin{figure}
\centering
\includegraphics[width=14cm,height=5cm]{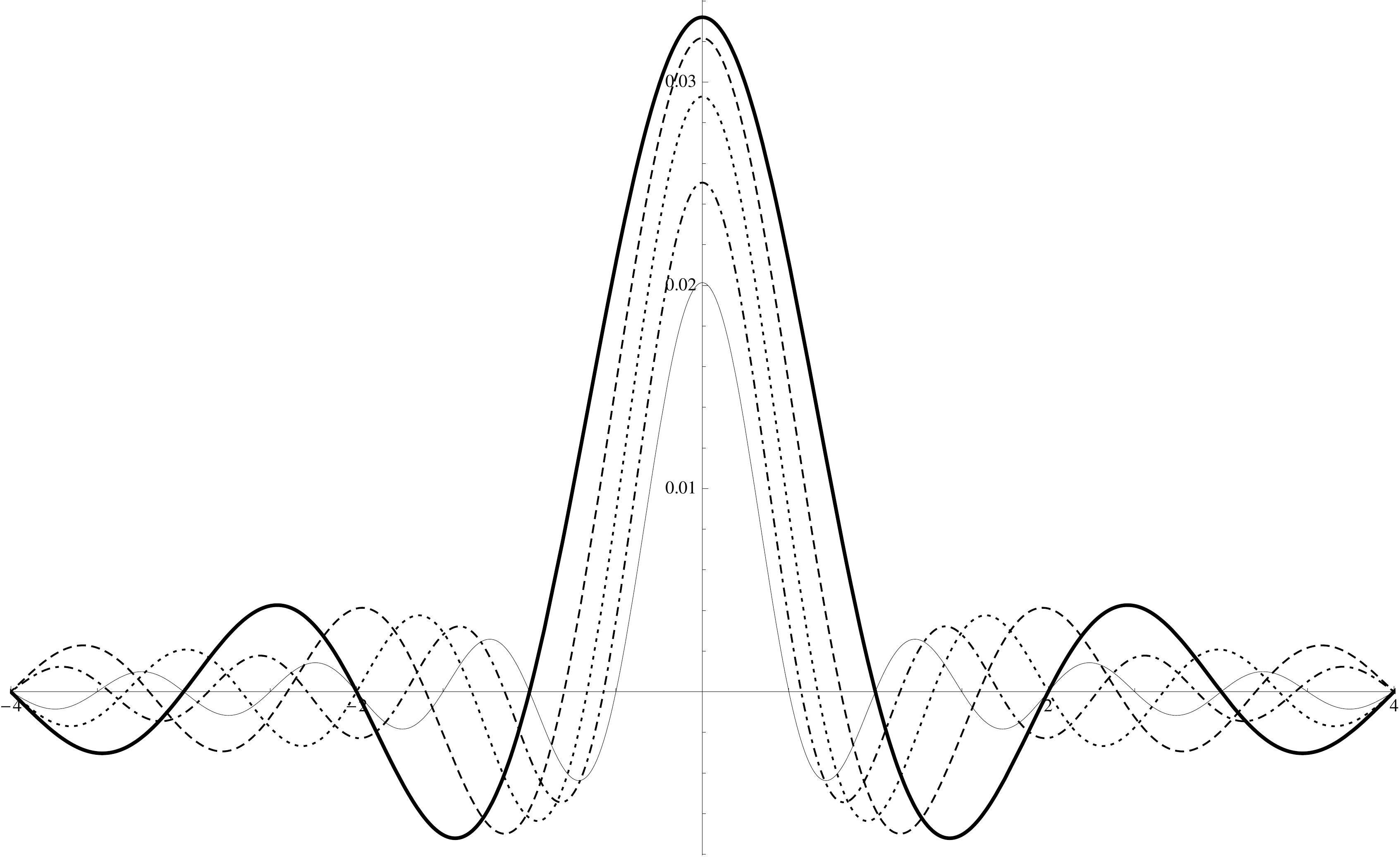}
\raisebox{1cm}{$\frac{r}{\Rh}$}
\caption{Modes of momentum number $k=k_{\rm c}$ (thick solid line), $k=(5/4)\,k_{\rm c}$ (dashed line),
$k=(3/2)\,k_{\rm c}$ (dotted line), $k=(7/4)\,k_{\rm c}$ (dash-dotted line)
and $k=2\,k_{\rm c}$ (thin solid line). The relative weight is determined according to Eq.~\eqref{psi_i}.
\label{modes}}
\end{figure}
\par\noindent

Let us start assuming a continuous distribution in momentum space of each of the $N$
scalar states given by half a Gaussian peaked around $k_{\rm c}$ (Fig.~\ref{modes}
displays a few modes above $k_{\rm c}$),
\be
\ket{\psi_{\rm S}^{(i)}}
=
\mathcal{N}_\gamma
\left(
\ket{k_{\rm c}}
+
\gamma_1
\int_{k_{\rm c}}^\infty
\frac{\sqrt{2}\,\d k_i}{\sqrt{\Delta_i\,\sqrt{\pi}}}\,
e^{-\frac{\hbar^2(k_i-k_{\rm c})^2}{2\,\Delta_i^2}}
\ket{k_i}
\right)
\ ,
\label{psi_i}
\ee
where $i=1,\ldots,N$, the ket $\ket{k}$ denotes the eigenmode of eigenvalue $k$,
and
\be
\mathcal{N}_\gamma
=
\left(1+\gamma_1^2\right)^{-1/2}
\ee
is a normalisation factor. 
Also, the width of the gaussians is $\Delta_i=m\simeq M/N\simeq \mpl/\sqrt{N}$, 
as calculated from the typical mode spatial size $k_{\rm c}^{-1}\sim \sqrt{N}\,\lp$,
and is the same for all particles.
This is, of course, a necessary simplification needed in order to simplify the calculations.
(More generally, one could assume a different width for each mode $k_i$.)
Since $m=\hbar\,k_{\rm c}$ and $E_i=\hbar\,k_i$, one can also write
\be
\ket{\psi_{\rm S}^{(i)}}
=
\mathcal{N}_\gamma
\left(
\ket{m}
+
\gamma_1
\int_{m}^\infty
\frac{\sqrt{2}\,\d E_i}{\sqrt{m\,\sqrt{\pi}}}\,
e^{-\frac{(E_i-m)^2}{2\,m^2}}
\ket{E_i}
\right)
\ .
\label{psiEi}
\ee
\par
The total wave-function will be given by the symmetrised product of $N$ such states,
and can be written by grouping equal powers of $\gamma_1$ as follows

\be
\ket{\Psi_{\rm N}}
&\!\!\simeq\!\!&
\frac{1}{N!}\,
\sum_{\{\sigma_i\}}^N
\left[
\bigotimes_{i=1}^N
\ket{m}
\right]
\nonumber
\\
&&
+
\frac{\gamma_1}{N!}
\left(\frac{2}{m\,\sqrt{\pi}}\right)^{1/2}
\,
\sum_{\{\sigma_i\}}^{N}
\left[
\bigotimes_{i=2}^{N}
\ket{m}
\otimes
\int_{m}^\infty
\d E_1\,
e^{-\frac{(E_1-m)^2}{2\,m^2}}
\ket{E_1}
\right]
\nonumber
\\
&&
+
\frac{\gamma_1^2}{N!}
\left(\frac{2}{m\,\sqrt{\pi}}\right)
\sum_{\{\sigma_i\}}^{N}
\left[
\bigotimes_{i=3}^{N}
\ket{m}
\otimes
\int_{m}^\infty
\d E_1\,
e^{-\frac{(E_1-m)^2}{2\,m^2}}
\ket{E_1}
\otimes
\int_{m}^\infty
\d E_2\,
e^{-\frac{(E_2-m)^2}{2\,m^2}}
\ket{E_2}
\right]
\nonumber
\\
&&
+
\ldots
\nonumber
\\
&&
+
\frac{\gamma_1^J}{N!}
\left(\frac{2}{m\,\sqrt{\pi}}\right)^{J/2}\,
\sum_{\{\sigma_i\}}^{N}
\left[
\bigotimes_{i=J+1}^{N}
\ket{m}
\,
\bigotimes_{j=1}^{J}
\int_{m}^\infty
\d E_j\,
e^{-\frac{(E_j-m)^2}{2\,m^2}}
\ket{E_j}
\right]
\nonumber
\\
&&
+
\ldots
\nonumber
\\
&&
+
\frac{\gamma_1^N}{N!}
\left(\frac{2}{m\,\sqrt{\pi}}\right)^{N/2}\,
\sum_{\{\sigma_i\}}^{N}
\left[
\bigotimes_{i=1}^{N}
\int_{m}^\infty
\d E_i\,
e^{-\frac{(E_i-m)^2}{2\,m^2}}
\ket{E_i}
\right]
\ ,
\label{NGausG}
\ee
where the power of $\gamma_1$ clearly equals the number of bosons in an
``excited'' mode with $k>k_{\rm c}$.
One can then identify two regimes, depending on the value of $\gamma_1$
(for further details, readers are directed to the Appendix~A in Ref.~\cite{Casadio:2014vja}). 
\par
For $\gamma_1\ll 1$, to leading order in $\gamma_1$, the spectral coefficient for
$E\ge M$ is given by the term corresponding to just one particle in the continuum,
\be
C(E\ge M)
\simeq
\mathcal{N}_\gamma
\left[
\delta_{E,M}
+
\gamma_1
\left(\frac{2}{m\,\sqrt{\pi}}\right)^{1/2}
e^{-\frac{(E-M)^2}{2\,m^2}}
\right]
\ ,
\ee
where $\delta_{A,B}$ is a Kronecker delta for the discrete part of the spectrum.
The width $m\sim \mpl/\sqrt{N}$ is the typical energy of Hawking quanta
emitted by a black hole of mass $M\simeq \sqrt{N}\,\mpl$.
The expectation value of the energy to next-to-leading order for large $N$ and small
$\gamma_1$ is
\be
\expec{E}
&\!\!\simeq\!\!&
\mathcal{N}_\gamma^2
\left(
M+
\int_M^\infty
E\,C^2(E)\,\d E
\right)
\nonumber
\\
&\!\!\simeq\!\!&
\sqrt{N}\,\mpl
\left(1+\frac{\gamma_1^2/\sqrt{\pi}}{1+\gamma_1^2}\,\frac{1}{N}
\right)
\nonumber
\\
&\!\!\simeq\!\!&
\sqrt{N}\,\mpl
\left(1+\frac{\gamma_1^2}{\sqrt{\pi}\,N}
\right)
\ ,
\ee
and its uncertainty 
\be
\Delta E
=
\sqrt{\expec{E^2}-\expec{E}^2}
\simeq
\frac{\gamma_1\,\mpl}{\sqrt{2\,N}}
\ .
\ee
One can also calculate the ratio
\be
\frac{\Delta E}{\expec{E}}
\simeq
\frac{\gamma_1}{\sqrt{2}\,N}
\ ,
\label{DEg}
\ee
where only the leading order in the large $N$ expansion was kept.
From the expression of the Schwarzschild radius $\rh=2\,\lp\,E/\mpl$,
one easily finds $\expec{\hat r_{\rm H}}\simeq \Rh$, with $\Rh$ given in
Eq.~\eqref{Rh}, and
\be
\frac{\Delta\rh}{\expec{\hat r_{\rm H}}}
\sim
\frac{1}{N}
\ ,
\ee
which vanishes rapidly for large $N$.
This case describes a macroscopic BEC black hole with (very) little
quantum hair, in agreement with Refs.~\cite{DvaliGomez}, thus overcoming
the problem of the large fluctuations~\eqref{Mmp} which appear in the case of
a single massive particle.
\par
This ``hair'' is expected to represent the Hawking radiation field,
and the connection with the thermal Hawking radiation
will become more clear in Section~\ref{ThBH}, where a different
form of the continuous spectrum will be employed
\subsection{Quantum hair with no black hole}

When $\gamma_1\gtrsim 1$ and $N\gg 1$, all of the $N$ particles are in the continuum,
while the ground state $\phi_{k_{\rm c}}$ is totally depleted.
The coefficient $\gamma_1^N$ and any overall factors can be omitted in this case,
and one finds 
\be
C(E\ge M)
\simeq
\int_m^\infty
\d E_1\cdots
\int_m^\infty
\d E_N\,
\exp\left\{-\sum_{i=1}^N\frac{(E_i-m)^2}{2\,m^2}\right\}
\,\delta\left(E-\sum_{i=1}^N E_i\right)
\ ,
\label{CE1}
\ee
along with $C(E<M)\simeq 0$.
Note that, in this case the mass $M=N\,m$ still represents the minimum energy of the system
corresponding to the ``ideal'' black hole with all the $N$ particles in the ground state $\ket{k_{\rm c}}$.
For $N=M/m\gg1$, this spectral function is estimated analytically in Ref.~\cite{Casadio:2014vja},
and is given by 
\be
C(E\ge M)
\simeq
\sqrt{\frac{2}{\pi\,m^3}}
\,(E-M)\,
e^{-\frac{(E-M)^2}{4\,m^2}}
\ ,
\label{CE>M}
\ee
which is peaked slightly above $E\simeq M=N\,m$, with a width
$\sqrt{2}\,\Delta_i\sim m$, so that the (normalised)
expectation value
\be
\expec{E}
\simeq
\int_M^\infty
E\,C^2(E)\,\d E
=
M+2\,\sqrt{\frac{2}{\pi}}\,m
\ ,
\ee
consistently for a system built out of continuous modes.
To be in the continuous part of the spectrum the energy of these modes must be (slightly) larger than $m$.
For $N\gg 1$, however, $\expec{E}=M\,[1+\mathcal{O}(N^{-1})]$, and
the energy quickly approaches the minimum value $M$.
This is also confirmed by the uncertainty
\be
\Delta E
\simeq
\sqrt{\frac{3\,\pi-8}{\pi}}\,m
\ ,
\ee
or $\Delta E\sim N^{-1/2}$.
\par
\begin{figure}[t]
\centering
\includegraphics[width=14cm]{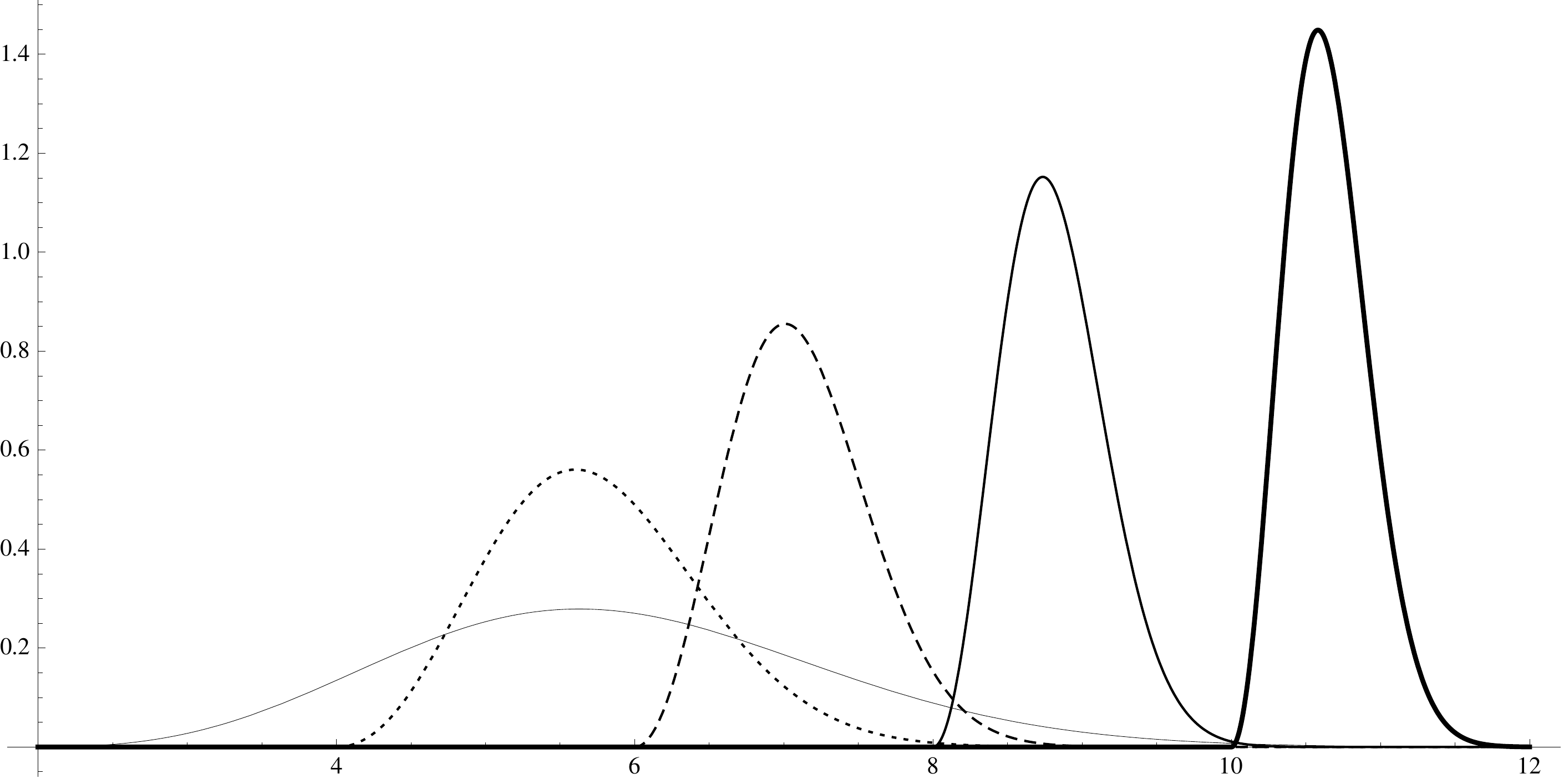}
\raisebox{0cm}{$\frac{\rh}{\lp}$}
\caption{Probability density for the gravitational radius $\rh$ for
$N=1$ ($\Rh=2\,\lp$; thin solid line),
$N=4$ ($\Rh=4\,\lp$; dotted line),
$N=9$ ($\Rh=6\,\lp$; dashed line),
$N=16$ ($\Rh=8\,\lp$; solid line)
and
$N=25$ ($\Rh=10\,\lp$; thick solid line).
The curves clearly become narrower the larger $N$. 
\label{plotRh}}
\end{figure}
The corresponding horizon wave-function is obtained by using $\rh=2\,\lp\,E/\mpl$,
and is approximately given by
\be
\psi_{\rm H}(\rh\ge 2\,\sqrt{N}\,\lp)
\simeq
\left(\rh-2\,\sqrt{N}\,\lp\right)
\,
e^{-\frac{\left(\rh-2\,\sqrt{N}\,\lp\right)^2}{16\,\lp^2/N}}
\ ,
\ee
and $\psi_{\rm H}(\rh< 2\,\sqrt{N}\,\lp)\simeq 0$.
The probability density of finding the gravitational radius between $\rh$ and $\rh+\d\rh$
is plotted in Fig.~\ref{plotRh} for different values of $N$.
The plot shows that for $N\sim 1$, the uncertainty in the size of the gravitational radius
is large, but it decreases very fast with increasing $N$.
The expectation value
\be
\expec{\hat r_{\rm H}}
\simeq
2\,\sqrt{N}\,\lp
\left(
1 
+\sqrt{\frac{2}{\pi}}\,\frac{2}{N}
\right)
=
\Rh
\left[
1 
+
\mathcal{O}(N^{-1})
\right]
\ ,
\ee
approaches (from above) the horizon radius of the ideal black hole,
$\Rh=2\,\sqrt{N}\,\lp$, for large $N$.
Since $\expec{\hat r_{\rm H}}>\Rh$, one can safely view it as a trapping surface,
whose uncertainty is proportional to the energy $m=\mpl/\sqrt{N}$, that is
\be
\frac{\Delta\rh}{\expec{\hat r_{\rm H}}}
=
\frac{\sqrt{\expec{\hat r_{\rm H}^2-\expec{\hat r_{\rm H}}^2}}}{\expec{\hat r_{\rm H}}}
\simeq
\frac{1}{N}
\ ,
\ee
which vanishes as fast as in the previous case for large $N$.
This is also expected in a proper semiclassical regime~\cite{DvaliGomez}.
\par
It needs to be emphasised that cases with $\gamma_1\not\ll 1$ do not describe a BEC black hole,
because most or all of the gravitons are in some excited mode with $k>k_{\rm c}$.
However, one can think that these states may play a role either at the threshold of black hole formation
(before the gravitons condense into the ground state $\ket{k_{\rm c}}$ and form a BEC)
or near the end of black hole evaporation.
\subsection{BEC with thermal quantum hair}
\label{ThBH}
\par\noindent

In Section~\ref{BHhair} it was shown that, for $\gamma_1\ll 1$, the quantum state of $N$ scalars is dominated by the configurations with just one boson in the continuum of excited states,
while the remaining $N-1$ gravitons form the BEC.
A Gaussian distribution was used to describe the continuous part of the spectrum,
and it was found that the spectral function has a typical width of the order of the Hawking
temperature,
\be
T_{\rm H}
=
\frac{\mpl^2}{4\,\pi\,M}
\simeq
\frac{\mpl}{\sqrt{N}}
\ ,
\ee
or $T_{\rm H}\simeq m$.
It is thus logical to ask what happens if the Gaussian distribution in Eq.~\eqref{psi_i} is replaced
by a thermal spectrum at the temperature $T_{\rm H}$.
\par
We hence start with a Planckian distribution~\footnote{Using a Boltzmann distribution for $N=1$
is also possible.
However, for $N>1$, it is affected by IR divergences.}
at the temperature $T_{\rm H}=m$ for the continuum part, namely
\be
\ket{\psi^{(i)}_{\rm S}}
&\!\!=\!\!&
\frac{\mathcal{N}_{\rm H}}{m^{3/2}}
\int_{m}^\infty
\d \omega_i\,
{\frac{(\omega_i-m)}{\exp\left\{\frac{\omega_i-m}{m}\right\}-1}\Ket{\omega_i}}
\nonumber
\\
&\!\!\equiv\!\!&
\mathcal{N}_{\rm H}
\int_{0}^\infty
\d \mathcal{E}_i\,G(\mathcal{E}_i)\,\ket{\mathcal{E}_i}
\ ,
\label{ThermalWF}
\ee 
where $\mathcal{N}_{\rm H}=\sqrt{3}/\sqrt{\pi^2-6\,\zeta(3)}\simeq 1.06$,
the dimensionless variables
\be
\mathcal{E}_i
=
\frac{\omega_i-m}{m}
\  ,
\label{dimless}
\ee
and the states $\ket{\mathcal{E}_i}=m^{1/2}\,\ket{\omega_i}$ were introduced, such that the identity
in the continuum can be written as
\be
\int_0^\infty
\d \mathcal{E}_i
\ket{\mathcal{E}_i}\bra{\mathcal{E}_i}
=
\int_m^\infty
\d \omega_i
\ket{\omega_i}\bra{\omega_i}
=
\mathbb{I}
\ .
\ee
The ``thermal hair'' is again assumed to appear due to the scatterings between the scalars inside the BEC,
therefore the parameter $\gamma_1$ should be related to the toy gravitons self-coupling $\alpha$.
In this exercise it is however kept as a free variable, so that one could relate it a posteriori
to known features of the Hawking radiation.
\par
Assuming that $\gamma_1\not=0$ yields a good enough approximation of the leading effects due to
these bosons self-interactions, the BEC can be treated as made of otherwise free scalars.
The total wave-function of the system of $N$ such bosons can again be written by collecting
powers of $\gamma_1$,

\be
\ket{\Psi_N}
&\!\!\simeq\!\!&
\frac{1}{N!}
\sum_{\{\sigma_i\}}^N
\left[
\bigotimes_{i=1}^N
\ket{m}
\right]
\nonumber
\\
&&
+
\gamma_1\,
\frac{\mathcal{N}_{\rm H}}{N!}\,
\sum_{\{\sigma_i\}}^{N}
\left[
\bigotimes_{i=2}^{N}
\ket{m}
\otimes
\int_{0}^\infty
\d \mathcal{E}_1\,
G(\mathcal{E}_1)
\ket{\mathcal{E}_1}
\right]
\nonumber
\\
&&
+
\gamma_1^2\,
\frac{\mathcal{N}_{\rm H}^2}{N!}
\sum_{\{\sigma_i\}}^{N}
\left[
\bigotimes_{i=3}^{N}
\ket{m}
\otimes
\int_{0}^\infty
\d \mathcal{E}_1\,
G(\mathcal{E}_1)
\ket{\mathcal{E}_1}
\otimes
\int_{0}^\infty
\d \mathcal{E}_2\,
G(\mathcal{E}_2)
\ket{\mathcal{E}_2}
\right]
\nonumber
\\
&&
+
\ldots
\nonumber
\\
&&
+
\gamma_1^J\,
\frac{\mathcal{N}_{\rm H}^{J}}{N!}
\sum_{\{\sigma_i\}}^{N}
\left[
\bigotimes_{i=J+1}^{N}
\ket{m}
\,
\bigotimes_{j=1}^{J}
\int_{0}^\infty
\d \mathcal{E}_j\,G(\mathcal{E}_j)
\ket{\mathcal{E}_j}
\right]
\nonumber
\\
&&
+
\ldots
\nonumber
\\
&&
+
\gamma_1^N\,
\frac{\mathcal{N}_{\rm H}^{N}}{N!}
\sum_{\{\sigma_i\}}^{N}
\left[
\bigotimes_{i=1}^{N}
\int_{0}^\infty
\d \mathcal{E}_i\,G(\mathcal{E}_i)
\ket{\mathcal{E}_i}
\right]
\ ,
\ee
where the overall normalisation constant of $1/(1+\gamma_1^2)^{N/2}$ was omitted for
the sake of simplicity.
\par
For $E=M=N\,m$, on neglecting an overall normalisation factor, one obviously has
\be
C(M)
\simeq
\frac{1}{N!}
\bra{M}
\sum_{\{\sigma_i\}}^N
\left[
\bigotimes_{i=1}^N
\ket{m}
\right]
=
1
\ .
\ee
For energies above the ground state, $E>M=N\,m$, the spectral coefficients are given by
\be
C(E>M)
=
\pro{E}{\Psi_N}
\ ,
\ee
and using the dimensionless variables~\eqref{dimless},
along with
\be
\mathcal{E}
=
\frac{E-M}{m}
\ ,
\ee
one finds
\be
C(\mathcal{E}>0)
&\!\!\simeq\!\!&
\gamma_1\,
\mathcal{N}_{\rm H}\,
G(\mathcal{E})
\nonumber
\\
&&
+
\gamma_1^2\,
\mathcal{N}_{\rm H}^2
\int_0^\infty
G(\mathcal{E}_1)\,
G(\mathcal{E}-\mathcal{E}_1)\,
\d \mathcal{E}_1\,
\nonumber
\\
&&
+\ldots
\nonumber
\\
&&
+
\gamma_1^N\,
{\mathcal{N}_{\rm H}^N}
\int_0^\infty
G(\mathcal{E}_1)\, \d \mathcal{E}_1
\times\dots\times
\int_0^\infty
G(\mathcal{E}_N)\, \d \mathcal{E}_{N}
\,
\delta\!\left(\mathcal{E}-\sum_{i=1}^{N}{\mathcal{E}_i}\right)
\nonumber
\\
&\!\!\equiv\!\!&
\sum_{n=1}^N{\gamma_1^n\,C_n(\mathcal{E})}
\ ,
\label{Cseries}
\ee
where all the coefficients in this expression can be written as
\be
C_n
=
{\mathcal{N}_{\rm H}^{n}}
\int_0^\infty
G(\mathcal{E}_1)\, \d \mathcal{E}_1
\times\dots\times
\int_0^\infty
G(\mathcal{E}_{n-1})\, \d \mathcal{E}_{n-1}\,
G\!\left(\mathcal{E}-\sum_{i=1}^{n-1}{\mathcal{E}_i}\right)
\ .
\label{BCoeff}
\ee
Each integral in $\mathcal{E}_i$ is peaked around $\mathcal{E}_i=0$,
so that one can approximate
\be
G\!\left(\mathcal{E}-\sum_{i=1}^{n-1}{\mathcal{E}_i}\right)
=
\frac{\mathcal{E}-\sum_{i=1}^{n-1}{\mathcal{E}_i}}
{\exp\left\{\mathcal{E}-\sum_{i=1}^{n-1}{\mathcal{E}_i}\right\}-1}
\simeq
\frac{\mathcal{E}}
{\exp\left\{\mathcal{E}\right\}-1}
=
G(\mathcal{E})
\ ,
\ee
for $2\le n\le N$.
Therefore
\be
C_n
&\!\!\simeq\!\!&
\mathcal{N}_{\rm H}
\left(\mathcal{N}_{\rm H}\,\frac{\pi^2}{6}\right)^{n-1}
G(\mathcal{E})
=
(1.75)^{n-1}\,\mathcal{N}_{\rm H}\,
G(\mathcal{E})
\ .
\ee
Upon rescaling 
\be
\gamma\simeq 0.57\,\sum_{j=1}^N\left(1.75\,\gamma_1\right)^j
\ ,
\label{Capprox}
\ee
and switching back to dimensionful variables, one finds 
\be
C(E>M)
&\!\!\simeq\!\!&
\gamma\,
\frac{\mathcal{N}_{\rm H}}{\sqrt{m}}\,
\frac{(E-M)/m}
{\exp\left\{(E-M)/m\right\}-1}
\ .
\label{C(E)ThermApprox}
\ee
This approximation was checked numerically to work extremely well for a wide
range of $N$.
The details of the numerical analysis can be found in the Appendix~B
of Ref.~\cite{Casadio:2014vja}.
\par
The result in Eq.~\eqref{C(E)ThermApprox} means that one can describe
the quantum state of the $N$-particle system as the (normalised) single-particle state
\be
\ket{\Psi_{\rm S}}
\simeq
\frac{\ket{M}+ \gamma\ket{\psi}}{\sqrt{1+\gamma^2}}
\ ,
\label{1PEF}
\ee
where
\be
\ket{\psi}
=
\frac{\mathcal{N}_{\rm H}}{\sqrt{m}}
\frac{(E-M)/m}{\mathrm{exp}\left\{ (E-M)/m \right\}-1} \ket{E}
\ .
\ee
Therefore, the BEC black hole effectively looks like one particle of very large mass $M=N\,m$
in a superposition of states with ``Planckian hair''.
To have most of the toy gravitons in the ground state, all that needs to be assumed is
$\gamma\ll 1$.
However, nothing prevents one from also considering the above approximate wave-function for
$\gamma\simeq 1$ or even larger.
\subsubsection{Partition function and entropy}
\label{Entropy}
\par\noindent

It is now straightforward to employ the effective wave-function~\eqref{1PEF}
to compute expectations values, such as
\be
\expec{\hat H}
\!\!&=&\!\!
\frac{1}{1+\gamma^2}
\left\{
M+ \gamma^2 \,\frac{\mathcal{N}^2_{\rm H}}{m}
\int_M^\infty \left[
\frac{(E-M)/m}{\mathrm{exp}\left\{(E-M)/m\right\}-1}
\right]^2 \, E \, \d E
\right\}
\nonumber
\\
\!\!&=&\!\!
\mpl\, \sqrt{N}
\left[
1 + \gamma^2\,\frac{\mathcal{N}^2_{\rm H}}{N}
\, 
\left(6\zeta(3)-\frac{\pi^4}{15}\right)
\right]
+ O(\gamma^4)
\ ,
\label{expE}
\ee
where the higher powers of the parameter $\gamma$ were discarded and the
approximation $(1+\gamma^2)^{-1} \simeq 1-\gamma^2$ was used. 
\par
Since
\be
\expec{\hat H^2}
&\!\!=\!\!&
\frac{1}{1+\gamma^2}
\left\{
M^2 + \gamma^2 \,\frac{ \mathcal{N}^2_{\rm H}}{m}
\int_M^\infty \left[
\frac{(E-M)/m}{\mathrm{exp}\left\{(E-M)/m\right\}-1}
\right]^2 \, E^2 \, \d E
\right\}
\nonumber
\\
&\!\!=\!\!&
\mpl^2 \, N
\left[
1 + 2 \, \frac{\gamma^2}{N} \, \mathcal{N}^2_{\rm H} 
\, \left( 6\,\zeta(3)-\frac{\pi^4}{15} \right)
+ \frac{4}{15} \, \frac{\gamma^2}{N^2} \, 
\mathcal{N}^2_{\rm H} \, \left(\pi^4-90\,\zeta(5)\right)
\right]
+O(\gamma^4)
\ ,
\ee
one can easily obtain the uncertainty
\be
\Delta E
=
\sqrt{\expec{\hat H^2}-\expec{\hat H}^2}
&=&
\gamma\,\frac{\mpl}{\sqrt{N}} \, \mathcal{N}_{\rm H} \,
\sqrt{\frac{4}{15} \, \left(\pi^4-90\zeta(5)\right)
-\mathcal{N}^2_{\rm H} \left(6\zeta(3)-\frac{\pi^4}{15}\right)^2} 
+ O(\gamma^2)
\nonumber
\\
&\simeq&
0.76\, \gamma\,\frac{\mpl}{\sqrt{N}}
\ .
\label{uncE}
\ee
In the large $N$ limit, one recovers the result~\cite{Casadio:2014vja}
\be
\frac{\Delta E}{\expec{\hat H}}
\sim
\frac{\gamma}{N} + O(\gamma^2)
\ .
\ee
\par
From a thermodynamical point of view, Eq.~\eqref{expE} can be used
to estimate the partition function of the system according to
\be
\expec{\hat H}
=
- \frac{\partial}{\partial \beta} \log{Z(\beta)}
\ ,
\ee
where $\beta=\Th^{-1}=m^{-1}\simeq {\sqrt{N}}/{\mpl}$.
One then finds
\be
\expec{\hat H(\beta)}
&\!\!=\!\!&
\mpl^2\, \beta
\left(
1 + \frac{\K\, \gamma^2}{\mpl^2\, \beta^2}
\right)
\nonumber
\\
&\!\!=\!\!&
\frac{\partial}{\partial \beta}
\left[
\frac{1}{2}\,\mpl^2\,\beta^2 + \K\,\gamma^2\, \log(k\,\beta)
\right]
\ ,
\ee
where $\K=\mathcal{N}^2_{\rm H}\left[6\zeta(3)-{\pi^4}/{15}\right] \simeq 0.81$,
and $k$ is an integration constant with dimensions of mass.
It is then straightforward to obtain
\be
\log{Z(\beta)}
=
-\frac{1}{2}\,\mpl^2\,\beta^2
-\K\,\gamma^2\, \log(\beta\,k)
\ ,
\ee
and, for $k=\mpl$, one finds
\be
Z(\beta)
=
\left(\mpl\,\beta\right)^{-\K\gamma^2}
\,
e^{-\frac{1}{2}\,\mpl^2 \beta^2}
\ ,
\ee
which contains two factors.
One of the factors has the form $1/\beta=\Th$ to some power, which looks like the thermal
wave-length of a particle, and is due to the contribution of the excited modes to the energy spectrum.
This is consistent with these modes propagating freely, since an external potential was not
considered in the model~\footnote{In other words, the grey-body factors for bosonic
Hawking quanta are approximated by one.}.
The exponential damping factor appears because of the bound states which are localised
within the classical Schwarzschild radius $\Rh$.
\par
The statistical canonical entropy can next be obtained from
\be
S(\beta)
=
\beta^2\,
\frac{\partial F(\beta)}{\partial \beta}
\ ,
\ee
where $F(\beta)=-(1/\beta)\,\log{Z}$ is the Helmoltz free energy .
It is straightforward to get
\be
S(\beta)
=
\frac{1}{2}\,\mpl^2\,\beta^2 
- \K\,\gamma^2 \log({\mpl\,\beta})
+ \K\,\gamma^2
\ ,
\label{entrobeta}
\ee
which is the usual Bekenstein-Hawking formula~\cite{Bekenstein:1997bt}
plus a logharitmic correction.
In fact, $\beta\simeq m^{-1}$ can be written as a function of the Schwarzschild radius,
\be
\beta
=
\frac{\Rh}{2\,\lp\,\mpl}
\ ,
\ee
the constant can be rescaled and one can define the area of the horizon as $\Ah=4\,\pi\,\Rh^2$,
hence yielding
\be
S
=
\frac{\Ah}{4\,\lp^2}
- \frac{\K}{2} \, \gamma^2 \,
\log\!\left({\frac{\Ah}{16\,\pi\,\lp^2}}\right)
\ .
\ee
The expression of the entropy depends on the ``collective'' parameter $\gamma$ defined in
Eq.~\eqref{Capprox}, rather than on the single-particle $\gamma_1$ which gives the probability
for each constituent to be a Hawking mode and which is determined by the details of the scatterings
that occur inside the BEC.
Therefore, even if the Hawking radiation were detectable, the details of the interactions inside the BEC
would hardly show directly. 
Still, $\gamma_1=0$ also implies $\gamma=0$.
The logarithmic correction switches off if the scatterings inside the BEC are negligible (and the Hawking
radiation is absent).
It needs to be emphasised that in this corpuscular model the number $N$ of bosons is conserved,
and this happens because both the black hole and its Hawking radiation are made of the same kind
of particles.
\par
The specific heat is given by
\be
C_V
&\!\!=\!\!&
\frac{\partial \expec{\hat H}}{\partial T}
\nonumber
\\
&\!\!\simeq\!\!&
-\mpl^2\,\beta^2
+\K\,\gamma^2
\ ,
\ee
which is negative for small $\gamma$ and large $\beta$ (or, equivalently, large
$N$), in agreement with the usual properties of the Hawking radiation,
but vanishes for $\beta\simeq \gamma/\mpl$, that is for $N_c\sim\gamma^2$.
Assuming for simplicity $1.75\,\gamma_1\sim 1$, so that $\gamma\sim N$
from Eq.~\eqref{Capprox}, one finds that the specific heat vanishes for
$N_c \sim 1$. As one would naively expect the Hawking radiation switches off
when there are no more particles to emit.
This is in agreement with the microcanonical picture of the Hawking evaporation
and with the estimate of the backreaction from the next section. 
One can also find more details on the microcanonical picture in Refs.~\cite{micro}
and references therein. 
\subsubsection{Horizon wave-function and backreaction}
\label{HWF}
\par\noindent

It is interesting to compute the HWF and investigate the existence of a trapping surface
in this case.
The main result in Ref.~\cite{Casadio:2014vja}, also presented in the previous sections,
is that the quantum $N$-particle states considered shows a horizon of size very close to
the expected classical value $\Rh$, with quantum fluctuations that die out for increasing $N$,
as one expects in the context of semiclassical gravity,
\be
\expec{\hat{r}_{\rm H}}
\simeq
\Rh
+{O}(N^{-1/2})
\ ,
\ee
\par
The horizon wave function can be written using the spectral coefficients~\eqref{C(E)ThermApprox} as
\be
\Ket{\Psi_{\rm H}}
=
\frac{\Ket{\Rh} + \gamma \Ket{\psi_{\rm H}}}
{\sqrt{1+\gamma^2}}
\ .
\label{hwf}
\ee
The state $\Ket{\Rh}$ defined so that
\be
\langle \Rh | \, \hat{r}_{\rm H} \, | \Rh \rangle
=
\Rh
\ ,
\ee
represents the discrete part of the spectrum and the states
\be
\psi_{\rm H}(\rh>\Rh)
\equiv
\pro{\rh}{\psi_{\rm H}}
=
\mathcal{N}_{\rm H}\,\sqrt{\frac{N}{4\,\pi\,\Rh^3}}\,
\frac{(\rh-\Rh)/(2\,\lp\,m/\mpl)}{\exp\left\{\frac{\rh-\Rh}{2\,\lp\,m/\mpl}\right\}-1}
\ ,
\ee
are related to the excited Hawking modes.
As usual $\Psi(\rh\leq\Rh)\simeq 0$ and the normalization is fixed by the scalar
product~\eqref{Hpro}.
The expectation value of the gravitational radius is found to be 
\be
\expec{\hat{r}_{\rm H}}
\!\!&=&\!\!
4\pi
\int_{\Rh}^\infty
\left| \Psi_{\rm H}(\rh) \right|^2 \, \rh^3 \, \d \rh
\nonumber
\\
\!\!&=&\!\!
\Rh
\left[
1 + \frac{3\,\gamma^2}{N} \, \mathcal{N}^2_{\rm H} 
\left(6\,\zeta(3)-\frac{\pi^4}{15}\right)
\right]
+ O(\gamma^4)
\ ,
\ee
in which $\Rh$ is given by Eq.~\eqref{RH}.
We thus see that 
\be
\expec{\hat{r}_{\rm H}} - \Rh
=
3\,\gamma^2 \, \frac{\Rh}{N} + O\left(\frac{1}{N^2}\right)
\ .
\ee
From
\be
\expec{\hat{r}^2_{\rm H}}
=
\Rh^2
\left[
1 + 4\,\gamma^2 \, \mathcal{N}^2_{\rm H}
\left(6\zeta(3)-\frac{\pi^4}{15}\right) \frac{1}{N} 
\right]
+ O\left(\frac{1}{N^2}\right)
\ ,
\ee
one finds
\be
\Delta\rh
=
\Rh \, \frac{\gamma\, \mathcal{N}_{\rm H}}{\sqrt{N}}
\, \sqrt{2\left(6\zeta(3)-\frac{\pi^4}{15}\right)} +O\left(\frac{1}{N^2}\right)
\ee
and, omitting a factor of order one,
\be
\frac{\Delta\rh}{\expec{\hat{r}_{\rm H}}}
\sim
\frac{\gamma}{N}+O(\gamma^2)
\ ,
\ee
which is the same as the result obtained in Section \ref{BHhair} (and also Ref.~\cite{Casadio:2014vja}).
\par
Note that $\expec{\hat{r}_{\rm H}}>\Rh$, even if only by a small amount for
large $N$, which is in agreement with the Hawking quanta adding a contribution
to the gravitational radius.
The toy gravitons are therefore bound within a sphere of radius slightly larger than the pure BEC
value $\Rh$.
This should make the typical energy of the reciprocal $2 \to 2$ scatterings a little smaller,
resulting in a slightly smaller depletion rate
\be
\Gamma
\!\!&\sim&\!\!
\frac{1}{\expec{\hat{r}_{\rm H}}}
\nonumber
\\
\!\!&\simeq&\!\!
\frac{1}{\sqrt{N}\,\lp}
\left[
1 - \frac{3\,\gamma^2}{N} \, \mathcal{N}^2_{\rm H} 
\left(6\,\zeta(3)-\frac{\pi^4}{15}\right)
\right]
\ .
\label{eq:GrateB}
\ee
Assuming the above relation still works for small $N$,
the flux would become zero for a number of quanta equal to
\be
N_c
\!\!&\simeq&\!\!
{3\,\gamma^2}\, \mathcal{N}^2_{\rm H} 
\left(6\,\zeta(3)-\frac{\pi^4}{15}\right)
\nonumber
\\
\!\!&\simeq&\!\!
2.4\,\gamma^2
\ .
\ee

Recalling Eq.~\eqref{Capprox}, and assuming
$1.75\,\gamma_1\lesssim 1$, so that $\gamma\lesssim N$, one obtains $N_c\gtrsim 1$,
which is the same as the flux vanishing for a finite number of constituents.
While the above approximations might not hold for small $N$ values, this result was also
obtained in the microcanonical description of the Hawking radiation: the emitted flux vanishes for 
vanishingly small black hole mass~\cite{micro}.
This is also in agreement with the thermodynamical analysis from Section~\ref{Entropy}, 
and the vanishing of the specific heat for $N=N_c\sim 1$.
However, $N_c$ depends on the collective parameter $\gamma$,
which is directly related with the single-particle $\gamma_1$ for small $N$.
%
%
%
%
%
\section{Conclusions}
\label{secC}
\par\noindent

Black holes represent some of the most interesting macroscopic objects
since they are the most likely candidates to provide a bridge between classical 
General Relativity, and a possibile quantum theory of gravity.
The aim of this review is to give the reader an introduction to the corpuscular model
of black holes introduced recently in Refs.~\cite{DvaliGomez} with emphasis on features
that can be more easily described using the HWF developed in
Refs.~\cite{Cfuzzy,Casadio:2013aua,qhoop}.
This model encompasses all of the features of a black hole as given by its many-body structure,
contrary to standard lore, while space-time geometry emerges as a consequence of the latter.
It was shown that the quantum state for a self-gravitating static massless
scalar field~\cite{dvaliCL} is approximately a BEC of the form conjectured in Ref.~\cite{DvaliGomez}.
The size of the horizon was then estimated by means of the HWF.
The result is in agreement with the classical picture of a Schwarzschild black hole for large $N$
(when the energy of each scalar is much smaller than $\mpl$, but the total energy is well above $\mpl$).
The uncertainty in the horizon's size is typically of the order of the energy of the expected Hawking quanta,
the latter being proportional to $1/N$ like it was claimed in Refs.~\cite{DvaliGomez}.
This result is in clear contrast with the case in which one considers a single particle of mass
$m\gg\mpl$ (like in Eq.~\eqref{Mmp}), for which a proper semiclassical behaviour cannot
be recovered.
This further supports the idea that black holes must be composite objects made of very light
constituents.
\par
Starting from the picture presented here, one may begin to conjecture what could occur
during the black hole formation via the gravitational collapse of stars.
The case considered in Section~\ref{star} is a simplistic model for a Newtonian lump of ordinary matter:
the source $J\sim\rho$ represents the star, with
$\Bra{g}\hat {\phi}\Ket{g}\simeq\phi_{\rm c}\sim V_{\rm N}$
that reproduces the outer Newtonian potential.
The energy contribution of the gravitons is in this case assumed to be negligible.
In Section~\ref{BH} the situation is just the opposite:
the contribution of any matter source is neglected, the (almost monochromatic) quantum state
is self-sustained and roughly confined inside a region of size given by its Schwarzschild radius.
This state also satisfies the relations~\eqref{eq:Max}, which have been known to hold for a self-gravitating
body near the threshold of black hole formation (see, for example,
Refs.~\cite{Ruffini:1969qy,Chavanis:2011cz}).
There is no time dependence in this treatment and the two configurations cannot be linked explicitly.
However, the two cases look like approximate descriptions of the beginning and (a possible) end-state
of a collapsing star. 
\par
For this to make sense, one can think about a phase transition for the graviton state around the time
when matter and gravitons have comparable weight.
Such an analysis might help determine if the state of BEC black hole is ever reached
(according to Refs.~\cite{DvaliGomez,flassig}, a black hole is precisely
the state at the quantum phase transition point),
or the star just approaches that asymptotically, or maybe avoids it.
One could tackle this starting from the Klein-Gordon
equation with both matter and ``graviton'' currents,
\be
\Box\phi(x)
=
q_{\rm M}\,J_{\rm M}(x)
+q_{\rm G}\,J_{\rm G}(x)
\ ,
\label{eqTot}
\ee
where $J_{\rm M}$ is the general matter current employed in Section~\ref{star}
and $J_{\rm G}$ the graviton current given in Eq.~\eqref{Jg}.
For the configuration corresponding to a star, one could treat the latter as a perturbation,
by formally expanding for small $q_{\rm G}$.
This approximation is expected to lead to a correction for the Newtonian potential near the star,
since the graviton current~\eqref{Jg} is formally equivalent to a mass term for the ``graviton'' $\phi$.
In the opposite regime, the contribution from ordinary matter becomes small,
and one could instead expand for small $q_{\rm M}$.
Such a correction should increase the fuzziness of the horizon outside the black hole.
A phase transition might however happen when neither sources
are negligible, so to capture its
main features one could not use perturbative arguments.
In any case, an order parameter must be identified.
\par
Close to the phase transition, but before black hole formation, one might speculate that the case
in which all of the bosons are in a slightly excited state above the BEC energy ($\gamma_1\gtrsim 1$)
hints at the physical processes that could occur. 
Since the case $\gamma_1\ll 1$ of Section~\ref{BHhair}, or the equivalent Hawking case
of Section~\ref{ThBH}, represent the quantum state of a black hole which already formed,
one might view the parameter $\gamma_1$ (or the collective analogue $\gamma$)
as an effective order parameter for the transition from the star to a black hole.
In this case, in a dynamical context, $\gamma$ should acquire
a time dependence, thus decreasing from values of order one or larger
to much smaller values along the collapse.
Therefore, the possible dependence of $\gamma$
on the physical variables usually employed to define the state of matter
along the collapse is worth investigating.
\par
One needs to emphasise that the Newtonian approximation was used for the special
relativistic scalar equation.
However, in order to describe black holes one has to use General Relativity.
In this respect, the only general relativistic result used was the condition for 
the existence of trapping surfaces for spherically symmetric systems which lies at the
foundations the horizon wave-function formalism (and the Generalized Uncertainty 
Principle that follows from it~\cite{Casadio:2013aua}).
Similarly ``fuzzy'' descriptions of a black hole's horizon were obtained by 
quantising spherically symmetric space-time metrics, 
which do not require any knowledge of the quantum state of the source 
(see~\cite{davidson} and~\cite{ramy} and references therein).
Investigations of collapsing thin shells~\cite{torres} or thick shells~\cite{brustein}, 
also point towards similar scenarios.
The present approach is more general because it allows one
to connect the causal structure of space-time, encoded by the
horizon wave-function $\psi_{\rm H}$, to the presence of a material source
in a state $\psi_{\rm S}$.
In order to study the time evolution of the system, a ``feedback'' from $\psi_{\rm H}$
into $\psi_{\rm S}$ must be introduced~\cite{Casadio:2014twa}.
\par
Regarding the quantum state representing an evaporating
black hole made of toy scalar gravitons, it was found that a Planckian distribution at the Hawking 
temperature for the depleted modes leads to the main known
features of the Hawking radiation, like the Bekenstein-Hawking entropy,
the negative specific heat and Hawking flux, for large black hole mass (equivalently, large $N$).
This $N$-particle state is also found to be collectively
described very well by a one-particle wave-function in energy space.
Using this approximation one can find the leading order corrections
to the energy of the system that give rise to a logarithmic contribution to the
entropy, which ensures a vanishing specific heat for $N$ of order one 
(when we expect the evaporation stops).
This qualitative result is also supported by estimating the
backreaction of the Hawking radiation onto the horizon size using the horizon
wave-function of the system.
The Hawking flux depends on the energy scale of the BEC~\cite{DvaliGomez},
which is related to the horizon radius, therefore the extra contribution to the
latter by the depleted scalars precisely reduces the emission.
When extrapolating to small values of $N$, this 
reduction will at some point stop the Hawking radiation completely.
This is exactly what one expects from the conservation of the total 
energy of the black hole system~\cite{micro}.
\par
Let us conclude by emphasising once more that the analysis presented here
does not explicitly consider the time evolution, but only a snapshot of the system
at a given instant of time.
However, for a relatively small Hawking flux, a reliable approximation of the time
evolution is Eq.~\eqref{dotM} for large mass and $N$, or by the correspondingly
modified expression that follows from Eq.~\eqref{eq:GrateB} in the limit of $N$
approaching one.
Another important point is that for very small $N$, the BEC model should in
principle reduce to the description of black holes as single particle states
investigated in Refs.~\cite{Casadio:2014twa} and~\cite{Casadio:2015rwa}. 
%
%
\section*{Acknowledgments}
\par\noindent
R.~C.~and A.~G.~are partly supported by INFN grant FLAG.



\begin{thebibliography}{999} 
%
\bibitem{OS}
J.R.~Oppenheimer and H.~Snyder,
Phys.\ Rev.\  {\bf 56} (1939) 455;
J.R.~Oppenheimer and G.M.~Volkoff,
Phys.\ Rev.\  {\bf 55} (1939) 374.
%
\bibitem{joshi}
P.S.~Joshi,
``Gravitational Collapse and Spacetime Singularities,''
Cambridge Monographs on Mathematical Physics
(Cambridge, 2007).
%
%
\bibitem{Bekenstein:2004eh}
J.D.~Bekenstein,
``Black holes: Physics and astrophysics. Stellar-mass, supermassive and primordial black holes,''
astro-ph/0407560.
%
\bibitem{Thorne:1972ji}
K.S.~Thorne,
``Nonspherical Gravitational Collapse: A Short Review,''
in {\em J.R.~Klauder, Magic Without Magic,\/} San Francisco (1972), 231.
%
\bibitem{payne}
P.D.~D'Eath and P.N.~Payne,
Phys.\ Rev.\ D {\bf 46} (1992) 658;
Phys.\ Rev.\ D {\bf 46} (1992) 675;
Phys.\ Rev.\ D {\bf 46} (1992) 694.
%
\bibitem{Senovilla:2007dw}
J.M.M.~Senovilla,
Europhys.\ Lett.\  {\bf 81} (2008) 20004.
%
%
\bibitem{acmo}
G.L.~Alberghi, R.~Casadio, O.~Micu and A.~Orlandi,
JHEP {\bf 1109} (2011) 023.
%
\bibitem{hsu}
S.D.H.~Hsu,
Phys.\ Lett.\ B {\bf 555} (2003) 92,
%
\bibitem{calmet}
X.~Calmet, D.~Fragkakis and N.~Gausmann,
Eur.\ Phys.\ J.\ C {\bf 71} (2011) 1781;
%
X.~Calmet, W.~Gong and S.D.H.~Hsu,
Phys.\ Lett.\ B {\bf 668} (2008) 20.
%
\bibitem{hawking}
S.W.~Hawking, Nature {\bf 248}, (1974) 30;
Comm. Math. Phys. {\bf 43}, (1975) 199.
\bibitem{DvaliGomez} 
G.~Dvali and C.~Gomez,
JCAP {\bf 01}, 023 (2014);
``Black Hole's Information Group'',
arXiv:1307.7630;
Eur.\ Phys.\ J.\ C {\bf 74}, 2752 (2014);
Phys.\ Lett.\ B {\bf 719}, 419 (2013);
Phys.\ Lett.\ B {\bf 716}, 240 (2012);
Fortsch.\ Phys.\  {\bf 61}, 742 (2013);
G.~Dvali, C.~Gomez and S.~Mukhanov,
``Black Hole Masses are Quantized,''
arXiv:1106.5894 [hep-ph].
%
\bibitem{Cfuzzy} 
R.~Casadio,
``Localised particles and fuzzy horizons: A tool for probing Quantum Black Holes,''
arXiv:1305.3195 [gr-qc].
%
\bibitem{Casadio:2013aua} 
R.~Casadio and F.~Scardigli,
Eur.\ Phys.\ J.\ C {\bf 74}, no. 1, (2014) 2685 
  %
\bibitem{qhoop} 
R.~Casadio, O.~Micu and F.~Scardigli,
Phys. Lett. {\bf B 732} (2014) 105.
%
\bibitem{Casadio:2014twa}
  R.~Casadio,
  Eur.\ Phys.\ J.\ C {\bf 75} (2015) 4,  160
  %
\bibitem{Casadio:2015rwa}
  R.~Casadio, O.~Micu and D.~Stojkovic,
  JHEP {\bf 1505} (2015) 096
%
\bibitem{Casadio:2015sda}
  R.~Casadio, O.~Micu and D.~Stojkovic,
  Phys.\ Lett.\ B {\bf 747} (2015) 68
%
\bibitem{scardigli}
M.~Maggiore,
Phys. Lett. B {\bf 319} (1993) 83;
A.~Kempf, G.~Mangano, R.B.~Mann,
Phys. Rev. D {\bf 52} (1995) 1108;
F.~Scardigli,
Phys.\ Lett.\ B {\bf 452} (1999) 39;
F.~Scardigli and R.~Casadio,
Class.\ Quant.\ Grav.\  {\bf 20} (2003) 3915;
Int.\ J.\ Mod.\ Phys.\ D {\bf 18} (2009) 319.
%
\bibitem{PNC}
P.~Nicolini,
Int.\ J.\ Mod.\ Phys.\ A {\bf 24} (2009) 1229.
%
\bibitem{Casadio:2014vja}
R.~Casadio, A.~Giugno, O.~Micu and A.~Orlandi,
Phys.\ Rev.\ D {\bf 90} (2014) 084040 .
%
\bibitem{Casadio:2015bna}
R.~Casadio, A.~Giugno and A.~Orlandi,
Phys.\ Rev.\ D {\bf 91} (2015) 124069.
%
\bibitem{Mueck}
W.~M\"uck,
Eur.\ Phys.\ J.\ {\bf C73} (2013) 2679;
W.~M�ck and G.~Pozzo,
JHEP {\bf 05}, (2014) 128.
\bibitem{Foit:2015wqa}
V.~Foit and N.~Wintergerst,
``Self-similar Evaporation and Collapse in the Quantum Portrait
of Black Holes'',
arXiv:1504.04384 [hep-th].
\bibitem{Hofmann}
S.~Hofmann and T.~Rug, 
``A Quantum Bound-State Description of Black Holes'',
arXiv:1403.3224 [hep-th].
\bibitem{Kuhnel}
F.~Kuhnel, 
Phys.\ Rev.\ D\ {\bf 90} (2014) 084024; 
F.~Kuhnel and B.~Sundborg, 
``Modified Bose-Einstein Condensate Black Holes in d Dimensions'',
arXiv:1401.6067 [hep-th]; 
F.~Kuhnel and B.~Sundborg, 
JHEP {\bf 1412}, (2104) 016; 
F.~Kuhnel and B.~Sundborg, 
Phys.\ Rev.\ D\ {\bf 90} (2014) 064025.
\bibitem{Bekenstein:1997bt}
J.~D.~Bekenstein,
``Quantum black holes as atoms,''
gr-qc/9710076.
%
\bibitem{Ruffini:1969qy}
R.~Ruffini and S.~Bonazzola,
Phys.\ Rev.\  {\bf 187} (1969) 1767.
%
\bibitem{kuhnelBaryons}
F.~Kuhnel and M.~Sandstad,
``Baryon number conservation in Bose-Einstein condensate black holes,''
arXiv:1506.08823 [gr-qc].
%
\bibitem{dvaliCL}
G.~Dvali, C.~Gomez and A.~Kehagias,
JHEP {\bf 1111} (2011) 070;
G.~Dvali, G.F.~Giudice, C.~Gomez and A.~Kehagias,
JHEP {\bf 1108} (2011) 108.
 %
\bibitem{Colpi:1986ye}
M.~Colpi, S.~L.~Shapiro and I.~Wasserman,
 Phys.\ Rev.\ Lett.\  {\bf 57} (1986) 2485;
%
  M.~Membrado, J.~Abad, A.~F.~Pacheco and J.~Sanudo,
  Phys.\ Rev.\ D {\bf 40} (1989) 2736;
%
J.~Balakrishna,
``A Numerical study of boson stars: Einstein equations with a matter source,''
arXiv:gr-qc/9906110;
%
T.~M.~Nieuwenhuizen,
Europhys.\ Lett.\  {\bf 83} (2008) 10008;
%
T.~M.~Nieuwenhuizen and V.~Spicka,
``Bose-Einstein condensed supermassive black holes:
A Case of renormalized quantum field theory in curved space-time,''
arXiv:0910.5377 [gr-qc];
%
\bibitem{Chavanis:2011cz}
P.-H.~Chavanis and T.~Harko,
Phys.\ Rev.\ D {\bf 86} (2012) 064011.
%
\bibitem{duff} 
M.~J.~Duff,
Phys.\ Rev.\ D {\bf 7}, 2317 (1973).
%
\bibitem{deser} 
S.~Deser,
Gen.\ Rel.\ Grav.\  {\bf 42}, 641 (2010).
%
\bibitem{Casadio:2013hja}
R.~Casadio and A.~Orlandi,
JHEP {\bf 1308} (2013) 025.
%
\bibitem{micro}
B.~Harms and Y.~Leblanc,
Phys.\ Rev.\ D {\bf 46}, (1992) 2334;
R.~Casadio and B.~Harms,
Phys.\ Rev.\ D {\bf 58}, (1998) 044014;
Entropy {\bf 13}, (2011) 502.
%
\bibitem{flassig}
D.~Flassig, A.~Pritzel, and N.~Wintergerst,
Phys.\ Rev.\ D {\bf87}, (2013) 084007;
\bibitem{davidson}
A.~Davidson and B.~Yellin,
Phys. Lett. B, (2014) 267.
%
\bibitem{ramy}
R.~Brustein,
Fortsch.\ Phys.\  {\bf 62} (2014) 255;
R.~Brustein and M.~Hadad,
Phys.\ Lett.\ B {\bf 718} (2012) 653.
%
\bibitem{torres}
R.~Torres and F.~Fayos,
Phys.\ Lett.\ B {\bf 733} (2014) 169;
R.~Torres,
Phys.\ Lett.\ B {\bf 733} (2014) 21.
%
\bibitem{brustein}
R.~Brustein and A.~J.~M.~Medved,
JHEP {\bf 1309} (2013) 015.
%
%
%
\end{thebibliography}
\end{document}